\documentclass[aps,prd,superscriptaddress,nofootinbib,amsmath,amsfonts,preprintnumbers,groupedaddress,10pt,english]{revtex4}
\usepackage{amsmath}
\usepackage{amssymb}
\usepackage{babel}
\usepackage{wrapfig}
\usepackage{cancel}

\usepackage{relsize,exscale}
\makeatletter

\usepackage{array,multirow,graphicx}
\usepackage{soul}
\usepackage{dcolumn}

\usepackage{newlfont}
\usepackage{bm}
\usepackage[colorlinks,citecolor=blue,urlcolor=blue,linkcolor=blue]{hyperref}
\usepackage[figtopcap]{subfigure}
\usepackage{color}
\usepackage{verbatim}

\def\be{\begin{align}}
\def\ee{\end{align}}
\def\bea{\begin{eqnarray}}
\def\eea{\end{eqnarray}}
\def\bal{\begin{align}}
\def\eal{\end{align}}

\usepackage{scalerel}
\usepackage{tikz}
\usetikzlibrary{svg.path}
\definecolor{orcidlogocol}{HTML}{A6CE39}
\tikzset{
 orcidlogo/.pic={
 \fill[orcidlogocol] svg{M256,128c0,70.7-57.3,128-128,128C57.3,256,0,198.7,0,128C0,57.3,57.3,0,128,0C198.7,0,256,57.3,256,128z};
 \fill[white] svg{M86.3,186.2H70.9V79.1h15.4v48.4V186.2z}
 svg{M108.9,79.1h41.6c39.6,0,57,28.3,57,53.6c0,27.5-21.5,53.6-56.8,53.6h-41.8V79.1z M124.3,172.4h24.5c34.9,0,42.9-26.5,42.9-39.7c0-21.5-13.7-39.7-43.7-39.7h-23.7V172.4z}
 svg{M88.7,56.8c0,5.5-4.5,10.1-10.1,10.1c-5.6,0-10.1-4.6-10.1-10.1c0-5.6,4.5-10.1,10.1-10.1C84.2,46.7,88.7,51.3,88.7,56.8z};}}
\newcommand\orcid[1]{\href{https://orcid.org/#1}{\mbox{\scalerel*{
\begin{tikzpicture}[yscale=-1,transform shape]
\pic{orcidlogo};
\end{tikzpicture}
}{|}}}}
\graphicspath{{./}{Figs/}}
\begin{document}

\date{\today}

\title{Stable black holes in lower dimensional $f(\mathbb{Q})$ non-metric gravity}

\author{Gamal G.L. Nashed}
\email{nashed@bue.edu.eg}
\affiliation {Centre for Theoretical Physics, The British University in Egypt, P.O. Box
43, El Sherouk City, Cairo 11837, Egypt}
\author{Salvatore Capozziello}
\email{capozziello@na.infn.it}
\affiliation{Dipartimento di Fisica E. Pancini, Universit a di Napoli Federico II
Complesso Universitario di Monte Sant Angelo, Edificio G, Via Cinthia, I-80126, Napoli, Italy,}
\affiliation{Scuola Superiore Meridionale, Largo S. Marcellino 10, I-80138, Napoli, Italy,}
\affiliation{Istituto Nazionale di Fisica Nucleare (INFN), Sez. di Napoli
Complesso Universitario di Monte Sant Angelo, Edificio G, Via Cinthia, I-80126, Napoli, Italy.}

\begin{abstract}
We investigate exact charged and uncharged black hole solutions in a (2+1)-dimensional spacetime within the framework of  quadratic form of $f(\mathbb{Q})$ symmetric teleparallel gravity, where $\mathbb{Q}$ is the non-metricity scalar. By adopting spherical  symmetry and considering both vanishing and non-vanishing electromagnetic fields, we derive new classes of black hole solutions and analyze their geometric and physical properties. The study demonstrates that the inclusion of  quadratic corrections in the gravitational Lagrangian significantly modifies the structure of  solutions, producing deviations from the standard BTZ geometry. Invariants such as curvature and non-metricity scalars are calculated to classify the singularity structure and spacetime behavior. Thermodynamic quantities, including Hawking temperature, entropy, and heat capacity, are computed, showing consistency with the first law of black hole thermodynamics. Furthermore, we examine the geodesic motion of test particles and derive the effective potential to explore the stability of photon orbits. A notable outcome is the identification of weaker black hole singularities in comparison to General Relativity, attributed to the non-metricity corrections. The possibility of multi-horizon configurations is also explored. This study provides a comprehensive analysis of the gravitational, thermodynamic, and dynamical features of lower-dimensional black holes in $f(\mathbb{Q})$ gravity and highlights their distinct characteristics with respect to General Relativity.
\end{abstract}


\maketitle

\section{Introduction}
In the last few decades, gravity in lower dimensions has emerged as a significant area of study, explored from various perspectives. Foundational contributions to this field began appearing in the 1980s and 1990s \cite{Deser:1982vy,Deser:1983tn,Jackiw:1984je,Brown:1988kg,brown1988lower,Mann:1991qp,Mann:1995eu}, and by the following decade, singular solutions were identified \cite{Banados:1992wn,Carlip:1995qv,Martinez:1999qi} that describe spacetimes characterized by properties such as mass, electric charge, angular momentum, and a cosmological constant. Although this theoretical framework lacks gravitons due to the absence of dynamical degrees of freedom, the curvature equations still yield nontrivial spacetime configurations. These solutions lead to meaningful dynamical implications and offer a more manageable setting compared to the complexities of four-dimensional gravity. For a comprehensive analysis of (2+1)-dimensional geometries, see \cite{Podolsky:2021gsa}.


Additional black hole solutions in three dimensions can be obtained by coupling gravity to various matter fields or by incorporating higher-curvature terms in the gravitational action. A notable example of the latter is the new massive gravity \cite{Bergshoeff:2009hq}, along with its generalizations \cite{Gullu:2010pc,Nashed:2023pxd,Sinha:2010ai,Paulos:2010ke,Bueno:2022lhf,Nashed:2001im}, which admit black holes distinct from the BTZ solution when specific values of the coupling constants are chosen. These geometries often display novel asymptotic behaviors. They are not locally maximally symmetric, and typically develop curvature singularities \cite{Bergshoeff:2009aq,Oliva:2009ip,Nashed:2021pah,Clement:2009gq,Yousaf:2023ixg,Nashed:2005kn,Alkac:2016xlr,Barnich:2015dvt,Yousaf:2025svv,ElHanafy:2014efn,
AyonBeato:2009nh,Gabadadze:2012xv,Bhatti:2025tac,Nashed:2011fz,Yousaf:2025pso,Ayon-Beato:2014wla,Fareghbal:2014kfa,Nam:2010dd,Gurses:2019wpb,Bravo-Gaete:2020ftn,Chernicoff:2024dll, Chernicoff:2024lkj,Nashed:2021pkc}.
New classes of black holes also emerge when electromagnetic and scalar fields are present. For instance, solutions have been found within the Einstein-Maxwell framework \cite{Clement:1993kc,Kamata:1995zu,Nashed:2021gkp,Martinez:1999qi, Hirschmann:1995he, Cataldo:1996yr, Dias:2002ps, Cataldo:2004uw, Cataldo:2002fh}, as well as in theories involving dilaton fields \cite{Chan:1994qa, Fernando:1999bh, Chen:1998sa,Koikawa:1997am,Edery:2020kof,Edery:2022crs,Karakasis:2022fep,Priyadarshinee:2023cmi} and Brans-Dicke-type scalar couplings \cite{Sa:1995vs,Dias:2001xt}. Both minimal and non-minimal scalar-gravity interactions have also been explored \cite{Martinez:1996gn,Henneaux:2002wm,Correa:2011dt,Zhao:2013isa,Tang:2019jkn,Karakasis:2021lnq,Baake:2020tgk,Nashed:2021ldz,Karakasis:2021ttn,Arias:2022jax,Desa:2022gtw,Cardenas:2022jtz,Karakasis:2023hni,Maluf:2024svb}, including setups derived from dimensional reductions of Lovelock gravity \cite{Hennigar:2020drx,Hennigar:2020fkv,Ma:2020ufk,Konoplya:2020ibi,Lu:2020iav}. These models often produce solutions with curvature singularities and logarithmic decay in some fields, though, in certain cases, the scalar sector remains smooth throughout the geometry. Moreover, nonlinear extensions of electrodynamics, such as Born-Infeld-like theories \cite{Cataldo:1999wr,Myung:2008kd,Mazharimousavi:2011nd,Mazharimousavi:2014vza,Hendi:2017mgb,Guerrero:2021avm,Gonzalez:2021vwp,Maluf:2022jjc,Sardeshpande:2024bnk}, have been shown to support regular black holes under particular choices of parameters and charges \cite{Cataldo:2000ns,He:2017ujy,Nashed:2009hn,HabibMazharimousavi:2011gh}. In this study, we are particularly interested in a charged  BTZ geometry using the quadratic form of non-metricity theory.  As far as we know that this issue has not been tackled before.

In view of this analysis, let us shortly discuss non-metric gravity.
In  Einstein General Relativity, spacetime is conventionally modeled in the framework of  Riemannian geometry, where the Levi-Civita connection governs the trajectories of particles and fields. This approach inherently assumes the absence of both torsion $(\mathbb{T})$ and non-metricity $(\mathbb{Q})$, considering curvature, derived from metric,  as the sole geometric attribute. Nonetheless, choices of connections on a manifold can yield distinct, yet physically equivalent formulations of gravity. Permitting the existence of torsion or non-metricity enables the construction of modified gravity theories. For example, setting both curvature and non-metricity to zero while allowing torsion results in the teleparallel formulation of General Relativity \cite{Buchdahl:1970ldb,Aldrovandi:2013wha,Bhatti:2024yzw,Nashed:2021ctg,Maluf:2012yn}. Conversely, a geometry with vanishing curvature and torsion but nonzero non-metricity leads to symmetric teleparallel gravity \cite{Adak:2008gd,Shirafuji:1996im,Mol:2014ooa,Nashed:2015pga,Jarv:2018bgs}, whose generalization is referred to as $f(\mathbb{Q})$ gravity, where the function $f$ is a smooth function of $\mathbb{Q}$. Specifically, gravitational dynamics is governed by the non-metricity scalar $\mathbb{Q}$  \cite{BeltranJimenez:2017tkd} and then isometries are restored only as a particular gauge choice. A key motivation for adopting $f(\mathbb{Q})$ gravity
 \cite{Conroy:2017yln,Latorre:2017uve,Hohmann:2018wxu,Nashed:2023uvk,Heisenberg:2023lru} is its potential to resolve issues in cosmology and astrophysics, offering an explanation for the universe accelerated expansion without invoking dark energy. In other words, this theory extends the standard  General Relativity  by introducing a gravitational action that is an arbitrary function of $\mathbb{Q}$, paving the way for novel insights into gravitational behavior.

Recently, there is considerable interest  in exploring  $f(\mathbb{Q})$ gravity. For instance, Harko et al. \cite{Harko:2018gxr} analyzed the interaction between matter and gravity by considering power-law and exponential forms of the theory. Lazkoz et al. \cite{Lazkoz:2019sjl} performed an observational evaluation of various $f(\mathbb{Q})$ models, using redshift parameterizations to test their potential as substitutes for the $\Lambda$CDM model in explaining the universe accelerated expansion. Mandal et al. \cite{Mandal:2020lyq} employed energy condition analyses to pinpoint physically acceptable models consistent with rapid cosmic expansion. In a study of stellar structures, Lin and Zhai \cite{Lin:2021uqa} examined how modifications in $f(\mathbb{Q})$ influence the mass of compact stars, observing that negative corrections enhance mass, whereas positive ones diminish it. Mandal and Sahoo \cite{Mandal:2021bpd} investigated the equation of state (EoS) and Hubble parameters using the Pantheon data, highlighting distinct behavior of $f(\mathbb{Q})$ models compared to the standard cold dark matter framework. Lymperis \cite{Lymperis:2022oyo} assessed the cosmological implications of $f(\mathbb{Q})$ through the lens of an effective dark energy component. Koussour et al. \cite{Koussour:2023ulc} explored how various cosmic parameters evolve within this theoretical model. In Ref. \cite{Bajardi:2023vcc}, the minisuperspace quantum cosmology for $f(\mathbb{Q})$ has been developed while a cosmographic reconstruction of viable $f(\mathbb{Q})$ models is discussed  in \cite{Capozziello:2022wgl}.
In Ref.\cite{Capozziello:2022tvv, Capozziello:2024lsz}, slow-roll inflation in $f(\mathbb{Q})$ gravity is discussed considering differences with respect to the analogue $f(R)$ models.
de Araujo and Fortes \cite{deAraujo:2024gnb} applied $f(\mathbb{Q})$  gravity to polytropic star models, studying the role of non-metricity both internally and externally. Gravitational waves in the $f(\mathbb{Q})$ context are derived in \cite{Capozziello:2024vix, Capozziello:2024jir} showing no substantial difference with respect to General Relativity, while an extra scalar mode is derived if boundary terms are considered into dynamics \cite{Capozziello:2024zij, Capozziello:2023vne}. Quantum information is taken into account in \cite{Capozziello:2024mxh} with the interesting feature that it is preserved in $f(\mathbb{Q})$ non-metric gravity cosmology. While previous studies on $f(Q)$ gravity (Refs.\cite{Harko:2018gxr}--\cite{Capozziello:2024mxh}) mainly focused on cosmological models, stellar structures, and the universe's accelerated expansion, our work extends the framework to lower-dimensional black hole physics. Unlike earlier analyses that explored large-scale or homogeneous configurations, we derive exact charged and uncharged $(2+1)$-dimensional black hole solutions within the quadratic form of $f(Q)$ gravity and examine their thermodynamic and dynamical stability. { This approach reveals new features, such as weaker singularities and multi--horizon configurations, not previously identified in non-metric gravity formulations. A concise summary of the novel features and their relation to BTZ solutions is presented in the Conclusions.}

The aim of the present study is to derive charged/non-charged $(2+1)$-dimensional black hole solutions within the quadratic form of $f(\mathbb{Q})$ and study the stability of the resulting solutions using different techniques. The structure  of the article is as follows: In Section \ref{f(Q)} we provide a summary   of  non-metric gravity and its extended form and also, derive the charged field equation of  $f(\mathbb{Q})$. In Section \ref{sec3}, we derive $(2+1)$-dimensional black hole solutions for  the quadratic form of $f(\mathbb{Q})$. We study different cases of the resulting differential equations. The inherent physics of the different solutions, derived in Section \ref{sec3}, is investigated in Section \ref{inv}. One of the interesting results of this section  is the exact solution that can be derived in the quadratic form of $f(\mathbb{Q})$ gravity using the  coincident gauge in contrast to the 4-dimensions. Among the results derived in Section \ref{inv}, one is related to the quadratic form of $f(\mathbb{Q})$: It represents a weaker singular black hole than its General Relativity counterpart. In Section \ref{S41}, we study the thermodynamics of the black holes  showing that all the solutions are consistent with the standard thermodynamics presented in  literature. In Section \ref{sec:44},  we study the geodesic motions and   potential of the  solutions. In particular,  we derive  the second derivative of effective potential for null geodesics   showing the stability of  photon orbits. In Section \ref{multi}, we study the possibility to  create black holes with multi horizons using the  values of  parameters characterizing the solutions. We conclude and discuss the main results in  Section \ref{conclusion}.

\section{A summary of  non-metric gravity and its extension $f(\mathbb{Q})$     } \label{f(Q)}

In  $f(\mathbb{Q})$ gravity, the non-metricity is
represented as follows \cite{Novello:2008ra}:
\begin{equation}\label{1a}
\mathbb{Q}_{\lambda\iota\vartheta}=-g_{\iota\vartheta,\lambda}+g_{\vartheta\sigma}
\hat{\Gamma}^{\sigma}_{\iota\lambda}
+g_{\sigma\iota}\hat{\Gamma}^{\sigma}_{\vartheta\lambda},
\end{equation}
where $g_{\iota\vartheta}$ is  the metric tensor  and $\hat{\Gamma}^{\lambda}_{\iota\vartheta}$ is
the affine connection, which can be decomposed into three different expressions \cite{Hehl:1976kj} as:
\begin{equation}\label{2a}
\hat{\Gamma}^{\lambda}_{\iota\vartheta}={\Gamma}^{\lambda}_{\iota\vartheta}
+\mathbb{\gamma}^{\lambda}_{\;\iota\vartheta}+\mathbb{L}^{\lambda}_{\;\iota\vartheta},
\end{equation}
where ${\Gamma}^{\lambda}_{\iota\vartheta}$ is the Levi-Civita connection  represented as:
\begin{equation}\label{3a}
\Gamma^{\lambda}_{\iota\vartheta}=\frac{1}{2}g^{\lambda\sigma}
(g_{\sigma\vartheta,\iota}+g_{\sigma\iota,\vartheta}-g_{\iota\vartheta,\sigma}),
\end{equation}
and $\mathbb{\gamma}^{\lambda}_{\;\iota\vartheta}$ is the contortion tensor  expressed as\footnote{For the symmetric tensor, we use the round parentheses and we define it as $\Phi^{(ab)}=\frac{1}{2}(\Phi^{ab}+ \Phi^{ba})$. For the skew symmetric tensor, we use the square brackets and we defined it as $\Phi^{[ab]}=\frac{1}{2}(\Phi^{ab}- \Phi^{ba})$.}:
\begin{equation}\label{4a}
\mathbb{\gamma}^{\lambda}_{\;\iota\vartheta}=\hat{\Gamma}^{\lambda}_{[\iota\vartheta]}
+g^{\lambda\sigma}g_{\iota\kappa}\hat{\Gamma}^{\kappa}_{[\vartheta\sigma]}
+g^{\lambda\sigma}g_{\vartheta\kappa}\hat{\Gamma}^{\kappa}_{[\iota\sigma]}.
\end{equation}
Here we define, $\mathbb{L}^{\lambda}_{\;\iota\vartheta}$ as the disformation tensor  which is  expressed as:
\begin{equation}\label{5a}
\mathbb{L}^{\lambda}_{\;\iota\vartheta}=\frac{1}{2}g^{\lambda\sigma}(\mathbb{Q}_{\vartheta\iota\sigma}
+\mathbb{Q}_{\iota\vartheta\sigma}-\mathbb{Q}_{\sigma\iota\vartheta}).
\end{equation}
We can define  the superpotential as
\begin{equation}\label{6a}
\mathbb{P}^{\lambda}_{\;\iota\vartheta}=-\frac{1}{2}\mathbb{L}^{\lambda}_{\;\iota\vartheta}
+\frac{1}{4}(\mathbb{Q}^{\lambda}-\tilde{\mathbb{Q}}^{\lambda})g_{\iota\vartheta}-
\frac{1}{4} \delta ^{\lambda}\;_{({\iota}}\mathbb{Q}_{\vartheta)},
\end{equation}
where $\mathbb{Q}^{\lambda}$ and $\tilde{\mathbb{Q}}^{\lambda}$ are defined as:
\begin{equation}\label{7a}
\mathbb{Q}_{\lambda}=\mathbb{Q}^{~\iota}_{\lambda~\iota},\quad
\tilde{\mathbb{Q}}_{\lambda}=\mathbb{Q}^{\iota}_{~\lambda\iota}.
\end{equation}
The non-metricity scalar is \cite{Nashed:2024jmw,Nashed:2024ush,Nashed:2025usa}
\begin{equation}\label{8a}
\mathbb{Q}=-\mathbb{Q}_{\lambda\iota\vartheta}\mathbb{P}^{\lambda\iota\vartheta}
=-\frac{1}{4}(-\mathbb{Q}^{\lambda\vartheta\upsilon}\mathbb{Q}_{\lambda\vartheta\upsilon}
+2\mathbb{Q}^{\lambda\vartheta\upsilon}\mathbb{Q}_{\upsilon\lambda\vartheta}-2\mathbb{Q}^{\upsilon}\tilde{\mathbb{Q}}_{\upsilon}+\mathbb{Q}^{\upsilon}\mathbb{Q}_{\upsilon}).
\end{equation}
Using the above definitions, we can write the  (2+1) action for $f(\mathbb{Q})$ gravity   incorporating both  matter
 $L_{m}$ and  electromagnetic
$L_{\mathbf{e}}$ Lagrangians, as \cite{BeltranJimenez:2017tkd}
\begin{equation}\label{9a}
S=\int\frac{1}{2}f(\mathbb{Q})
\sqrt{-g}d^{3}x+\int(L_{m}+L_{\mathbf{e}}) \sqrt{-g}d^{3}x,
\end{equation}
where $ L_{\mathbf{e}}$ is the Lagrangian of the Maxwell field tensor  which is defined as
\begin{equation}\label{10a}
L_{\mathbf{e}}=-\frac{1}{16\pi} F_{\mu\nu} F^{\mu\nu}.
\end{equation}
where $ F_{\mu\nu}$ represents the three-potential electromagnetic field strength which is defined as
$F_{\mu\nu}=\varPhi_{\mu,\nu} - \varPhi_{\nu,\mu} $, with $\varPhi_{\mu}$ being the electromagnetic potential.  { In this study, we assume the electromagnetic four-potential to take the form
$\Phi_{\mu} = [\varphi(r),\,0,\,0,\,0]|$ where $\varphi(r)$ represents the electrostatic potential, which depends solely on the radial coordinate}. The  field equations of
$f(\mathbb{Q})$ gravity are:
\begin{equation}\label{11a}
\frac{-2}{\sqrt{-g}}\nabla_{\lambda}(f_{\mathbb{Q}}\sqrt{-g}
\mathbb{P}^{\lambda}_{\;\iota\vartheta})-\frac{1}{2}f
g_{\iota\vartheta}-f_{\mathbb{Q}}
(P_{\iota\varpi\varsigma}\mathbb{Q}_{\vartheta}^{~\varpi\varsigma}-2\mathbb{Q}^{\varpi\varsigma}_{~~~\iota}
P_{\varpi\varsigma\vartheta})=T_{\iota\vartheta}+\mathbb{E}_{\iota\vartheta}.
\end{equation}
The matter energy-momentum tensor is defined as:
\begin{equation}\label{29}
T_{\iota\vartheta} \equiv \frac{-2}{\sqrt{-g}} \frac{\delta
(\sqrt{-g} L_{M})}{\delta g^{\iota\vartheta}},
\end{equation}
with $f_{\mathbb{Q}}$ being  the derivative w.r.t.  $\mathbb{Q}$.
The energy-momentum tensor of  electromagnetic field considers is defined  on the curvature of spacetime. It is expressed as\footnote{In this study,  the matter  energy-momentum tensor, $T_{\iota\vartheta}$ can be equal zero because we are interested in  charged/non-charged vacuum solutions}:
\begin{equation}\label{12a}
\mathbb{E}_{\iota\vartheta} = \frac{1}{4\pi} \bigg( F^{\nu}_\iota
F_{\vartheta\nu} - \frac{1}{4} g_{\iota\vartheta} F_{\mu\nu}
F^{\mu\nu} \bigg).
\end{equation}
\section{Spherically symmetric black holes in quadratic  $f(\mathbb{Q})$ gravity}\label{sec3}
A spherically symmetric  metric is given by:
\begin{equation}\label{1}
ds^{2}=-\mu(r)dt^{2}+\frac{dr^{2}}{\nu(r)}+{ r^2}d\phi^2,
\end{equation}
where $\mu(r)$  and $\nu(r)$ are two arbitrary functions of the radial coordinates.  In $f(\mathbb{Q})$ gravity, the Einstein-Maxwell field equations are expressed as follows
\begin{eqnarray}\label{2}
0&=&\frac{f}{2}+\frac{2\nu }{r}f_{\mathbb{Q}\mathbb{Q}}\mathbb{Q}^{\prime}-
\frac{f_{\mathbb{Q}}}r\bigg[
r\mathbb{Q}-\nu'\bigg]-\frac{2\nu \varphi'^2}{\mu},
\\\label{3}
0&=&\frac{f}{2}-\frac{f_{\mathbb{Q}}}{r\mu}\bigg(
r \mu \mathbb{Q}-\nu \mu'\bigg)-\frac{2\nu \varphi'^2}{\mu},
\\ 0 &=&{\frac {2\,f  \mu^{2}-2\,  f_{\mathbb{Q}}
 \,\mathbb{Q}\mu^{2
}+f_{\mathbb{Q}}\, \nu'
 \mu' \mu  +2\, f_{\mathbb{Q}}\,\nu  \mu'' \mu  -f_{\mathbb{Q}}\,
\nu   \mu'^{2}+4\, \varphi'^{
2}\mu   \nu  +2\,f_{\mathbb{Q}\mathbb{Q}}\,\nu  \mu'
 \mathbb{Q} \mu  }{4 \mu^{2}}}
.\label{4}
\end{eqnarray}
The non-metricity scalar is given as
\begin{equation}\label{5}
\mathbb{Q}=-\frac{\mu' \nu}{r\mu},
\end{equation}
where prime represents derivative w.r.t.  $r$. The quadratic form of  $f(\mathbb{Q})$ theory is
\begin{equation}\label{6}
f(\mathbb{Q})=\mathbb{Q} +\frac{1}{2}\alpha\mathbb{Q}^2-2\Lambda
\end{equation}
with $\alpha$ is a non-zero dimensional quantity that has the dimension of $lenght^{2}$, and $\Lambda$  the cosmological constant.  In order to recover General Relativity\footnote{Precisely the Symmetric Teleparallel Equivalent of General Relativity (STEGR).}, the dimensional quantity $\alpha$ has to vanish, i.e., $\alpha=0$.   Substituting Eqs.\eqref{5} and \eqref{6} into \eqref{2}
through \eqref{4}, we obtain
\begin{align}\label{7}
&0=\frac{2\nu_1\left(2\alpha\nu_1 \mu-3r\alpha \nu'_1 \mu-{r}^{2} \right) \mu'-4\alpha\mu\nu_1{}^{2} \mu'' r-4 \varphi'^{2}\nu_1 {r}^{3}-3r\alpha \mu'^{2}\nu_1{}^{2}-2{r}^{2} \nu'_1 \mu-4\,\Lambda{r}^{3}}{4{r}^{3}}
,\\
\label{8}
&0=\frac {-2\, \mu' \nu_1 \,r-3\,
\alpha\, \mu'^{2}\nu^{2}-4\,\Lambda \,{r}^{2}-4\,{r}^{2} \varphi'^{2}\nu}{4{r}^{2}}, \\\label{9}
&0=\frac {-2\,r\nu_1\, \left( 2\, \mu'\alpha\,\nu_1 +r \right) \mu''+4\,{
r}^{2} \ \varphi'^{2}\nu_1
-\alpha\,\nu_1 \, \left(3r\nu' -\nu_1  \right)  \mu'^{2}- \mu'  \nu'_1 {r}^{2}-4\,
\Lambda\,{r}^{2}}{4{r}^{2}}\,,
\end{align}
where we have assumed that $\nu(r)=\mu(r)\nu_1(r)$.
Now we are going to  derive different solutions of the differential equation  \eqref{7}--\eqref{9} using different constraints.
\subsection{Uncharged black  hole with  $\alpha=0$ }\label{sol1}
In the case when $\varphi'=0$ and $\nu_1=1$,  the differential equation Eqs. \eqref{7}--\eqref{9} yields:\\
\begin{align}
\mu(r)={ -} \Lambda r^2-m, \qquad \nu_1=1,
\end{align}
which is the well-know solution of BTZ of General Relativity \cite{Banados:1992wn}.
\subsection{Uncharged black hole with $\alpha\neq 0$}\label{sol2}
For $\varphi'=0$ and $\nu_1=1$,  Eqs. \eqref{7}--\eqref{9} gives:\\
\begin{align}
\mu(r)=-\frac{(1\pm\sqrt{1-12\alpha \Lambda}) r^2}{6\alpha}-m, \qquad \nu_1=1.
\end{align}
The above solution shows that the dimensional quantity $\alpha$ cannot be equal zero. However,  assuming $\alpha=\frac{1}{12\Lambda}$,  we get:
\begin{align}
\mu(r)=-\Lambda_1 r^2-m\,, \quad \mbox{where} \quad \Lambda_1=2\Lambda\,,
\end{align}
which is also the BTZ solution.
\subsection{Charged black hole with  $\alpha=0$}\label{sol33}
In the case  $\varphi'=0$ and $\nu_1=1$,  Eqs. \eqref{7}--\eqref{9} yields:\\
\begin{align}\label{sol3}
\mu(r)=-\Lambda r^2-m-2c_2{}^2\ln \left( \frac{r}{r_0}\right),  \qquad \varphi=c_1+c_2\ln \left( \frac{r}{r_0}\right),\qquad \nu_1=1,
\end{align}
 where $r_0$ is some length scale. The above solution  is the charged BTZ solution \cite{Hendi:2012zz}.

\subsection{Charged black hole with $\alpha\neq0$}\label{sol4}
In the case  $\varphi'\neq0$ and $\alpha\neq0$,  Eqs. \eqref{7}--\eqref{9} gives:\\
\begin{align}\label{gsol}
&\mu(r)=\int \!{\frac { r \left[ 12\,\alpha\,{r}^{4}
\Lambda-{r}^{4}+12\,{r}^{2}{c_2}^{2}\alpha+\{4\,\alpha\,{c_2}^{2}-{r}^{2}+4\,{r}^{2}\alpha\,\Lambda\}\,\Upsilon(r)\right]r }{\alpha\, {\nu_1}(r) \left({r}^{2}-12{r}^{2}\alpha \Lambda-12\alpha{c_2}^{2}+2\Upsilon(r) \right) }}{dr}-m\,,  \nonumber\\
& \varphi=c_1+c_2\ln \left( \frac{r}{r_0}\right),\qquad \nu_1=\exp\int \frac {-8 \left(r^2-12r^2\alpha \Lambda-12c_2^2+2\Upsilon(r)\right) c_2^{2}\alpha}{ \left( 8\alpha {r}^{4}\Lambda-{r}^{4}+96{r}^{2}c_2^{2}\Lambda{\alpha}^{2
}+48{\alpha}^{2}c_2^{4}+8{r}^{2}c_2^{2}\alpha+48
{\alpha}^{2}{r}^{4}{\Lambda}^{2} \right) r}{dr},\nonumber\\
&\mbox{where} \qquad  \Upsilon(r)=\sqrt {
{r}^{2} \left(r^2- 12\,{r}^{2}\alpha\,\Lambda-12\,\alpha\,{
c_2}^{2} \right) }.
\end{align}
{ One can verify that, in the limit $\alpha \to 0$, Eq.~(\ref{gsol}) reduces smoothly to Eq.~(\ref{sol3}), recovering the charged BTZ black hole solution and when both $\alpha=0$ and $c_2=0$ we recover the well know BTZ solution. In the next section, we are going to study the  physical properties of  solutions presented in this section.}

\section{Metric spacetime and invariants }\label{inv}
Now we are going to construct the metric and invariants for the two cases presented in subsections \ref{sol2} and \ref{sol4}.
\subsection{Metric and invariants of the case $\alpha\neq 0$ and $\varphi=0$}
\begin{figure}
\centering
\subfigure[~The behavior of the metric given by Eqs. \eqref{met1}]{\label{fig:R}\includegraphics[scale=0.24]{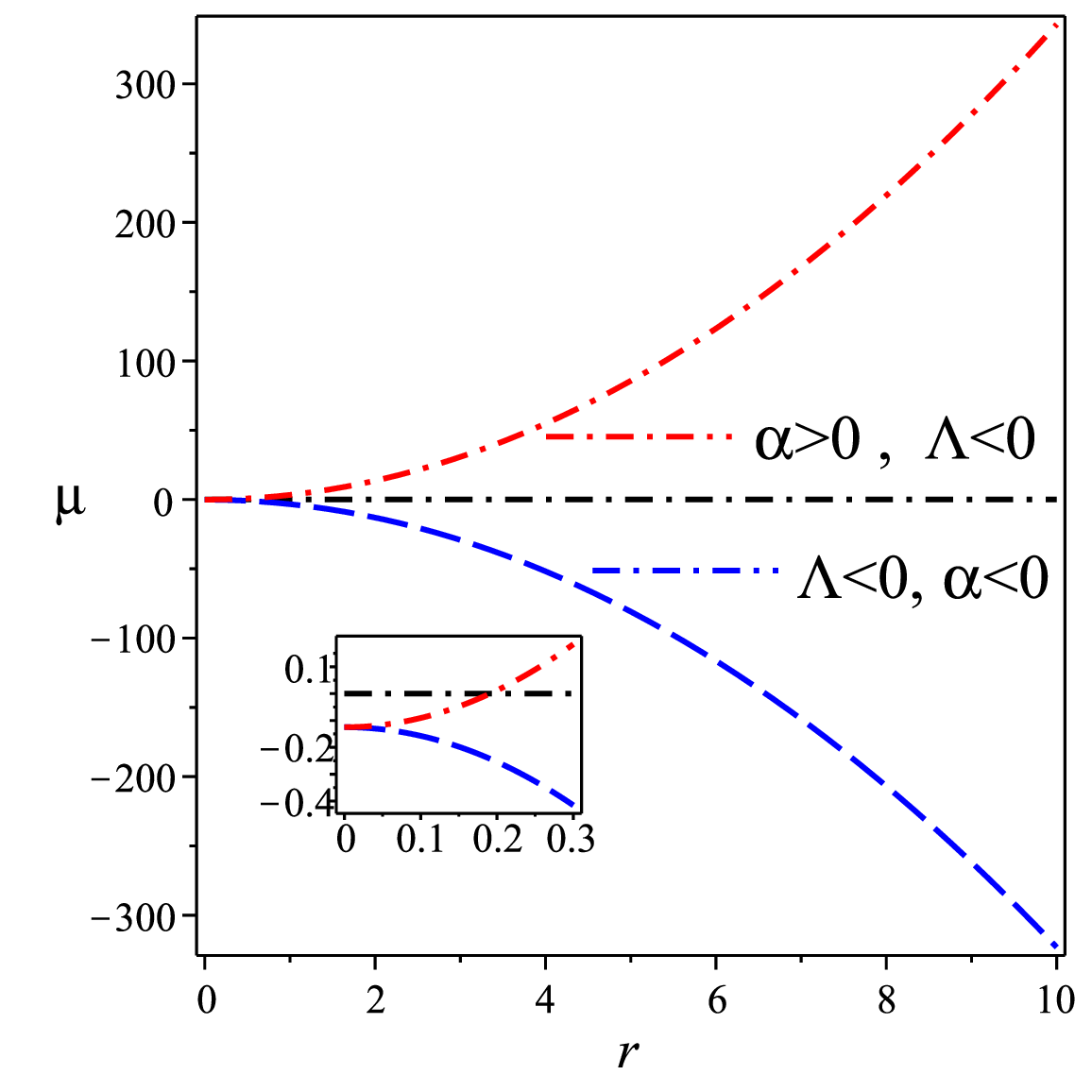}}
\subfigure[~The behavior of the metric given by Eqs. \eqref{2t}]{\label{fig:fr1}\includegraphics[scale=0.24]{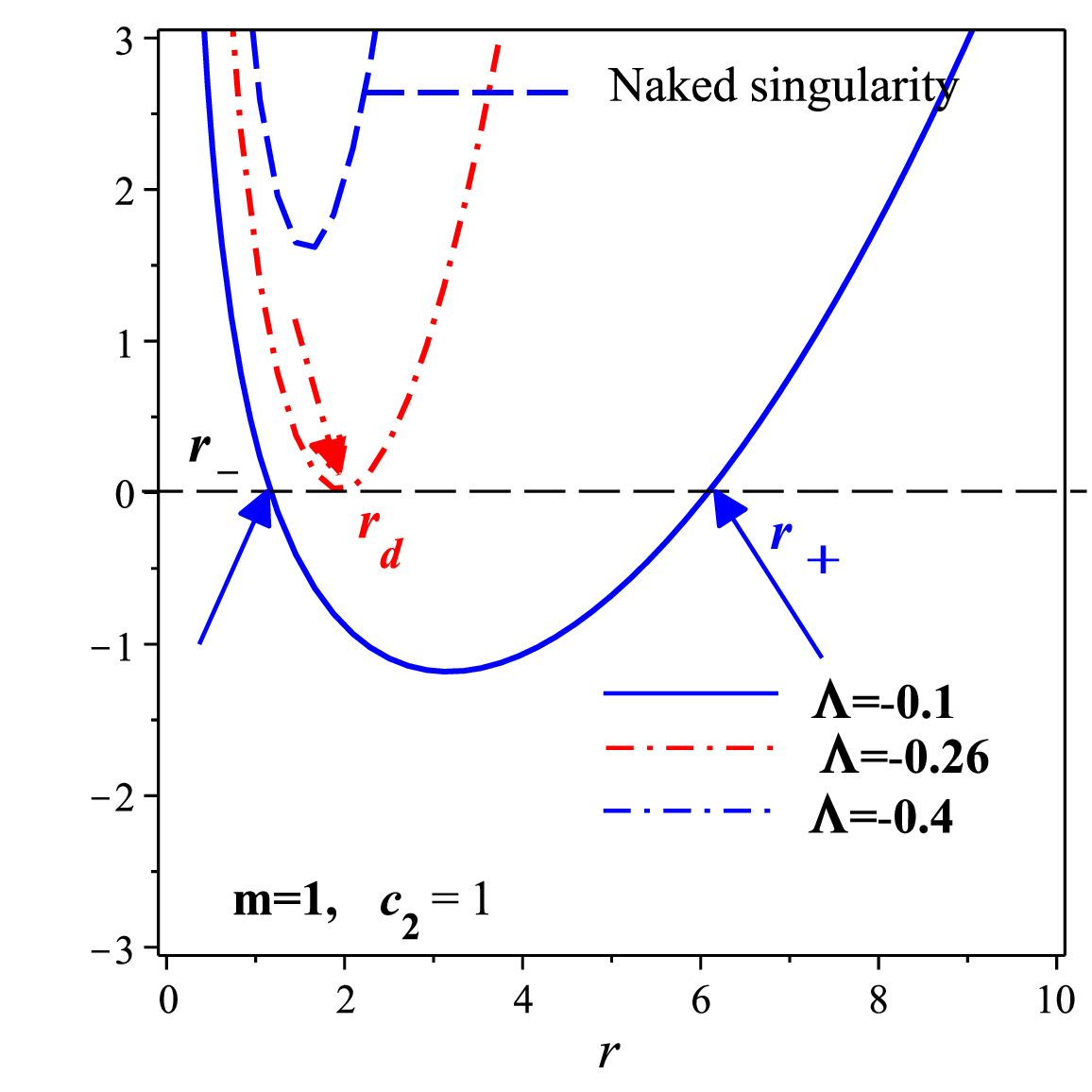}}
\subfigure[~The behavior of the metric given by Eqs. \eqref{met2}]{\label{fig:fr}\includegraphics[scale=0.24]{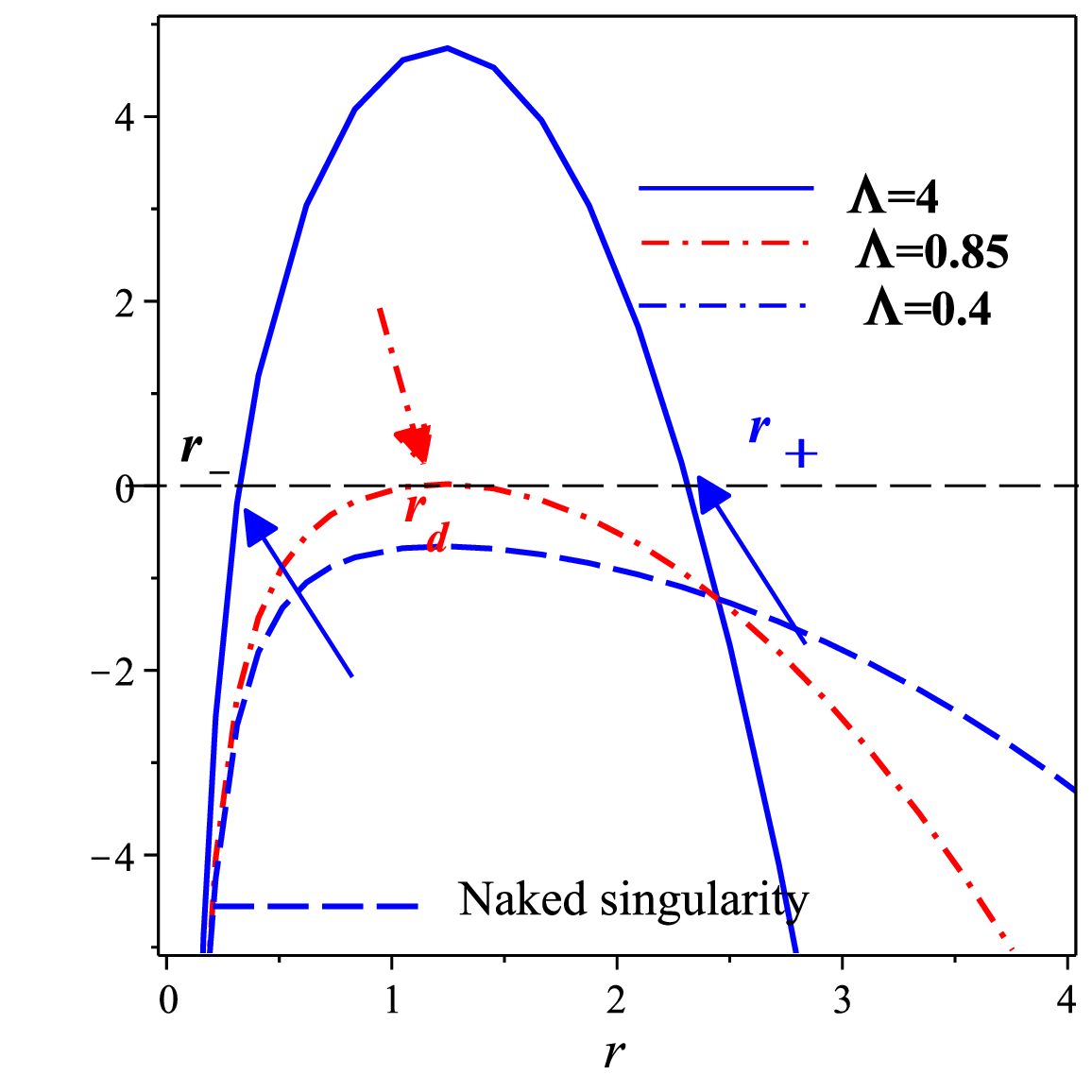}}
\caption{{~Radial behavior of the metric function for different black hole solutions.
Panel \subref{fig:R} shows the uncharged solution with $\alpha\neq0$ and $\phi=0$, given by Eq.~(\ref{met1}),
which effectively reproduces a BTZ-type geometry with a rescaled cosmological constant.
Different curves correspond to different signs of $\alpha$ and $\Lambda$, illustrating
how the quadratic $f(\mathbb{Q})$ correction modifies the effective (A)dS structure.
Panel \subref{fig:fr1} displays the charged solution for the special branch $\alpha=1/(12\Lambda)$,
given by Eq.~\eqref{2t}, where the presence of electric charge leads to horizon formation
and possible naked singularities depending on parameter choices.
Panel \subref{fig:fr} shows the charged generic branch $\alpha\neq1/(12\Lambda)$, described by Eq.~\eqref{met2},
which exhibits a non-BTZ asymptotic behavior dominated by quadratic non-metricity effects.
In all panels, deviations from the General Relativity (BTZ) case arise solely from the
quadratic $f(\mathbb{Q})$ corrections.}}
\label{Fig:1}
\end{figure}

In the case $\alpha\neq 0$ and $\varphi=0$, we get the metric in the form:
\begin{align}\label{met1}
ds^{2}=-\left[-\frac{(1\pm\sqrt{1-12\alpha \Lambda}) r^2}{6\alpha}-m\right]dt^{2}+\frac{dr^{2}}{-\frac{(1\pm\sqrt{1-12\alpha \Lambda}) r^2}{6\alpha}-m}+{ r^2}d\phi^2,
\end{align}
which is the BTZ solution. The above metric shows that we can create the AdS/dS spacetimes according to the sign of the dimensional quantity $\alpha$. The behavior of the metric \eqref{met1} is shown in Fig.~\ref{Fig:1} \subref{fig:R} for $\alpha>0$ and $\Lambda<0$ and for $\alpha<0$ and $\Lambda<0$. If we take the case  $\alpha>0$ and $\Lambda>0$,  and the case $\alpha<0$ and $\Lambda>0$,  the curves will reverse i.e., the blue curve becomes red one and the red becomes blue.
 The  invariants of the line-element \eqref{met1} yields:
\begin{align}\label{inv22}
&R_{\mu \nu \alpha \beta}R^{\mu \nu \alpha \beta}=R_{\mu \nu}R^{\mu \nu}=\pm \Lambda_{eff}^2\,, \qquad R=\mathbb{Q}=\pm\Lambda_{eff}\,,\nonumber\\
&{\mathbb{Q}}_{\mu \nu \alpha}{\mathbb{Q}}^{\mu \nu \alpha}=2\,{\frac {{\Lambda_{eff}}^{2}{r}^{4}-4\,\Lambda_{eff}\,{r}^{2}-m\,\alpha+12\,{-m}^{2}{\alpha}^{2}}{\alpha\, \left( 6-\Lambda_{eff}\,{r}^{2}\,-m\,\alpha\, \right) {r}^{2}}}\approx-{\frac {2\Lambda_{eff}}{\alpha}}+{4\frac {m}
{{r}^{2}}}-{48\frac {\alpha\,{m}^{2}}{{\Lambda_{eff}}\,{r}^{4}}}\nonumber\\
&{\mathbb{Q}}_{\mu}{\mathbb{Q}}^{\mu}=-\frac{\Lambda_{eff}}{3\alpha}-\frac{m}{r^2},\qquad \tilde{\mathbb{Q}}_\mu \tilde{\mathbb{Q}}^\mu={\frac {{{\Lambda_{eff}}}^{2}{r}^{2}}{6\alpha\, \left( \Lambda_{eff}\,{r}^{2}+6\,m\alpha\right) }}\,,
\end{align}
where $\Lambda_{eff}=(1+\sqrt{1-12\alpha \Lambda})$. All the above invariants  coincide with those of General Relativity. So, in the non-charged the solution, we get the coincidence with the General Relativity case.

\subsection{Metric and invariants of case $\alpha\neq 0$ and $\varphi\neq0$}
For $\alpha\neq 0$ and $\varphi\neq0$, we are going to discuss two cases separately: The case for $\alpha=\frac{1}{12\Lambda}$ and the case  $\alpha\neq \frac{1}{12\Lambda}$.\\ \\

\centerline{\underline{\textbf {The case $\alpha=\frac{1}{12\Lambda}$}}} \vspace{0.3cm}
In this case, the metric has  the form:
\begin{align}\label{met2222}
&ds^{2}=-\mu(r) dt^{2}+\frac{dr^{2}}{\mu(r)\nu_1(r)}+{ r^2}d\phi^2, \qquad \qquad \varphi=c_1+c_2\ln \left( \frac{r}{r_0}\right),
\end{align}
where $\mu$ and $\nu_1$ are give by:
\begin{align}\label{2t}
\mu=\int \frac{\!4 \left( 2\,{r}^{2} \left( |\Lambda|\right) ^{3 /2}\pm3\,c_2\,\Lambda\,r+{c_2}^{2}\sqrt {|\Lambda|} \right)\left( 4\,\Lambda\,{r}^{2}+{c_2}^{2} \right) }{ \nu_1 {c_3}{r}^{2} \left( 2\,r\sqrt {|\Lambda|}\pm c_2 \right)}{dr}-m\,, \qquad \qquad \nu_1=\frac{c_3 r^2 e^{\pm {2\arctan \left( 2\,{ \frac {\sqrt {|\Lambda|}r}{c_2}} \right)}}}{  4r^2\Lambda+ c_2^2 }\,.
\end{align}
{ It is important to mention that the integrands in  Eq.~\eqref{2t} contains a true singularity which occurs at $r=0$ which is consistent with the behavior of the invariants listed below. This singularity  is generally independent of $c_2$ $c_3$, $\nu_1$ and $\Lambda$ as long as these constants are finite and non-zero.
However, we have horizons when $r=\mp\frac{c_2}{2\sqrt{|\Lambda|}}$ which is an event horizons.}
The behavior of the metric given by Eq.\eqref{2t} is shown in Fig.~\ref{Fig:1} \subref{fig:fr1}.
Eq. \eqref{met2222} asymptotically behaves as:
\begin{align}\label{met222}
&ds^{2}\approx -\left[8\Lambda r^2-12c_2^2\Lambda \ln\left( \frac{r}{r_0}\right)-m+\frac{4c_2^3\sqrt{|\Lambda|}}{r}\right]dt^2+\left[\Psi_1 {r}^{3}+\Psi_2\,r^2+\Psi_3r +\Psi_4-\frac{1}{4c_2\sqrt{|\Lambda|}r}\right]dr^2+r^2d\phi^2\,,
\end{align}

where $\Psi_i, i=1\cdots 4$ are lengthy function listed in the Appendix A:

The  invariants of the line-element \eqref{met2} yields:
\begin{align}\label{inv22}
&R_{\mu \nu \alpha \beta}R^{\mu \nu \alpha \beta}=R_{\mu \nu}R^{\mu \nu}\approx 48\,\Lambda^2-64\,{\frac {c_2\,{|\Lambda|}^{3/2}}{r}}+{\cal O}\left(\frac{1}{r^2}\right)
\quad R\approx 12\,\Lambda-8{\frac {c_2\,{|\Lambda|}^{1/2}}{r}}+{\cal O}\left(\frac{1}{r^2}\right)
\nonumber\\
&{\mathbb{Q}}=4\Lambda-\frac{4c_2\sqrt{|\Lambda|}}r+{\cal O}\left(\frac{1}{r^2}\right), \qquad {\mathbb{Q}}_{\mu \nu \alpha}{\mathbb{Q}}^{\mu \nu \alpha}={\mathbb{Q}}_{\mu}{\mathbb{Q}}^{\mu}=\tilde{\mathbb{Q}}_\mu \tilde{\mathbb{Q}}^\mu \approx-24\,\Lambda+16\,{\frac {c_2\,{|\Lambda|}^{1/2}}{r}}+{\cal O}\left(\frac{1}{r^2}\right)
\end{align}
{ Now let us discuss the properties of the above invariants and compared these results  with the case $\alpha=0$ that given by Eq.~\eqref{inva000}. For the large $r$ branch under consideration we obtained the values of invariants as given by Eq.~\eqref{inv22}.

In $(2{+}1)$ dimensions the Weyl tensor vanishes and the quadratic invariants obey the exact relation
\begin{equation}\label{3Didentity}
R_{\mu\nu\alpha\beta}R^{\mu\nu\alpha\beta}
= 4\,R_{\mu\nu}R^{\mu\nu} - R^2.
\end{equation}
If, asymptotically, $R_{\mu\nu\alpha\beta}R^{\mu\nu\alpha\beta}=R_{\mu\nu}R^{\mu\nu}$ holds to leading orders, then \eqref{3Didentity}
implies $3\,R_{\mu\nu\alpha\beta}R^{\mu\nu\alpha\beta}=R^2$.
With $R\simeq 12\Lambda+\cdots$ this yields
$R_{\mu\nu\alpha\beta}R^{\mu\nu\alpha\beta}\simeq (12^2/3)\Lambda^2=48\Lambda^2$,
in agreement with the first term in \eqref{inv22}.
Expanding $R^2$ further also produces an $O(1/r)$ piece proportional to $\text{sgn}(\Lambda)\,c_2|\Lambda|^{1/2}$, which is consistent with the $1/r$ term in \eqref{inv22}.

From \eqref{inv22} one reads $R\simeq 12\Lambda$ and $\mathbb{Q}\simeq 4\Lambda$, i.e.
\begin{equation}
R \;\simeq\; 3\,\mathbb{Q}\qquad (r\to\infty),
\end{equation}
showing that the curvature and non-metricity scalars approach \emph{proportional} constants fixed by $\Lambda$ in this branch.
This reflects an asymptotically constant-curvature background (locally (A)dS$_3$) modified by non-metricity effects.

Unlike the BTZ case (where all quadratic invariants are strictly constant), \eqref{inv22} contains universal $O(1/r)$ corrections,
These $1/r$ tails signal \emph{long-range hair} associated with the parameter $c_2$ and are a hallmark of a non-Einstein (higher-curvature / non-metricity) branch or of weakened (Brown-Henneaux-type) boundary conditions.
The overall sign of the $1/r$ coefficients follows $\mathrm{sgn}(\Lambda)$; the notation with $|\Lambda|$ in \eqref{inv22} makes this explicit.


\centerline{\underline{\textbf {The case $\alpha\neq\frac{1}{12\Lambda}$}}}
In this case,  we get the metric of the form:
\begin{align}\label{met2}
&ds^{2}=-\mu(r) dt^{2}+\frac{dr^{2}}{\mu(r)\nu_1(r)}+{ r^2}d\phi^2, \qquad \qquad \varphi=c_1+c_2\ln \left( \frac{r}{r_0}\right),
\end{align}
where $\mu$ and $\nu_1$ are defined in Eq.~\eqref{gsol}.  Eq. \eqref{met2} behaves asymptotically as:
\begin{align}\label{met22}
&ds^{2}_{ r\to \infty}\approx -\left[\frac{1}{12{\alpha}^{3}{c_2}^{4}} \left( {\alpha}^{2}{\Lambda}^{2}-\frac{\alpha\Lambda}6+{\frac {1}{144}} \right)\sqrt {-3\alpha\,{c_2 }^{2}}{r}^{3}-\frac{1}2\,{\frac { \left( \alpha\,\Lambda+\frac{1}{36} \right) {r}^{2}}{{\alpha}^{2}{c_2}^{ 2}}}+\frac{1}{12}{\frac { \left( 12\,{\alpha}^{2}{c_2}^{2}\Lambda-\alpha\,{c_2}^{2} \right) \sqrt {-3\alpha\,c_2^{2}}r}{{\alpha}^{3}{c_2}^{4}}}\right.\nonumber\\
&\left.+\frac{1}{2}{\frac {2\,c_2^{2}m\,{\alpha}^{2}+2\,\alpha\,{c_2}^{2}\ln \left( \frac{r}{r_0}\right) }{{\alpha}^{2}{c_2}^{2}}}-\frac{2}{3}{\frac {\sqrt {-3\alpha\,{c_2}^{2}}}{r\alpha}}\right]dt^2+\left[\Psi_5 {r}^{3}+\Psi_6\,r^2+\Psi_7r +\Psi_8-\frac{\sqrt{3|\alpha|}}{2c_2r}\right]dr^2+r^2d\phi^2\,,
\end{align}
where $\Psi_i, i=5\cdots 8$ are defined in Appendix A.

We show the behavior of the metric \eqref{met2} in Fig.~\ref{Fig:1} \subref{fig:fr}. { As is evident from Eq.~\eqref{met22}, the leading term of both the temporal and spatial components of the metric behaves as  $r^3$. This asymptotic behavior differs from the well-known BTZ solution, which approaches
$r^2$
 at large $r$. The deviation arises from the contribution of the higher-curvature corrections encoded in the $\alpha Q^2$ term. These corrections effectively modify the cosmological background, leading to an enhanced power-law growth of the metric components. Physically, this indicates that the spacetime is no longer asymptotically (A)dS in the Einstein sense but belongs to a non-Einstein branch dominated by quadratic curvature (or non-metricity) effects. Such high-curvature terms alter the effective cosmological constant and thus modify the large-scale structure of the geometry.}

Before going on, let us   calculate the invariants of the above metric for $\alpha=0$.  They are:
\begin{align}\label{inva000}
&R_{\mu \nu \alpha \beta}R^{\mu \nu \alpha \beta}=R_{\mu \nu}R^{\mu \nu}\approx12\Lambda^2+\frac{8c_2^2\Lambda}{r^2}+\frac{12c_2^4}{r^4}\,, \qquad \qquad R=\mathbb{Q}\approx 6\Lambda+\frac{2c_2^2}{r^2}\,,\nonumber\\
&{\mathbb{Q}}_{\mu \nu \alpha}{\mathbb{Q}}^{\mu \nu \alpha}={\mathbb{Q}}_{\mu}{\mathbb{Q}}^{\mu}=\tilde{\mathbb{Q}}_\mu \tilde{\mathbb{Q}}^\mu\approx-12\Lambda+\frac{8c_2^2ln\left( \frac{r}{r_0}\right)-4m-16c_2^2}{r^2}.
\end{align}
Eq. \eqref{inva000} shows that the leading term for all the invariants is  ${\cal O}\left(\frac{1}{r^2}\right)$.
The  invariants of the line-element \eqref{met2} yields:
\begin{align}\label{inva00}
&R_{\mu \nu \alpha \beta}R^{\mu \nu \alpha \beta}=R_{\mu \nu}R^{\mu \nu}\approx12\Lambda^2+\frac{\xi_1}{r^2}+\frac{\xi_2}{r^4}\,, \qquad \qquad R=\mathbb{Q}\approx 6\Lambda+\frac{\xi_3}{r^2}\,,\nonumber\\
&{\mathbb{Q}}_{\mu \nu \alpha}{\mathbb{Q}}^{\mu \nu \alpha}={\mathbb{Q}}_{\mu}{\mathbb{Q}}^{\mu}=\tilde{\mathbb{Q}}_\mu \tilde{\mathbb{Q}}^\mu\approx-12\Lambda+\frac{\xi_4}{r^2}.
\end{align}
where $\xi_i, i=1\cdots 4$ are constants that depend on $c_2$ and $\Lambda$. All the above invariants  coincide with those of charged BTZ as $\alpha=0$. So,  in general, the asymptotic behavior of the invariants either with $\alpha=0$ or with $\alpha\neq 0$ are the same.

{ To characterize the singularity strength, we adopt the standard criterion based on the radial scaling of curvature and non-metricity invariants, supplemented by the physical interpretation provided by the Tipler and Krolak conditions. In this framework, a singularity is considered weaker when the divergence of the relevant invariants occurs with a lower radial power than in the corresponding General Relativity (BTZ) case, indicating milder tidal effects on infalling observers.

For the quadratic $f{\mathbb{Q}}$ model, this behavior arises in specific non-General Relativity branches and parameter ranges, where the integrals associated with the Tipler and Krolak criteria remain finite or diverge more slowly than in General Relativity. By contrast, the branch continuously connected to General Relativity reproduces the standard BTZ-type singularity structure. Hence, the singularity softening is a branch-dependent effect induced by the quadratic $f(Q)$ corrections.}

\section{Thermodynamics}\label{S41}
In this section, we first obtain the thermodynamic quantities
of the black hole solutions like the temperature and entropy, and
after we examine the validity of the first law of thermodynamics.
From the  Hawking-Bekenstein relation for the black hole temperature
 on the event horizon $r_+$ (the outermost one),
we can calculate this quantity on our black hole solutions.
\subsection{The case $\alpha\neq \frac{1}{12\Lambda}$}
In the case $\alpha\neq \frac{1}{12\Lambda}$, the Hawking temperature relation  is the follow through the definition of the surface
gravity ($\kappa$):
\begin{figure}
\centering
\subfigure[~The behavior of temperature given by Eq. (\ref{kGR})]{\label{fig:temp1}\includegraphics[scale=0.21]{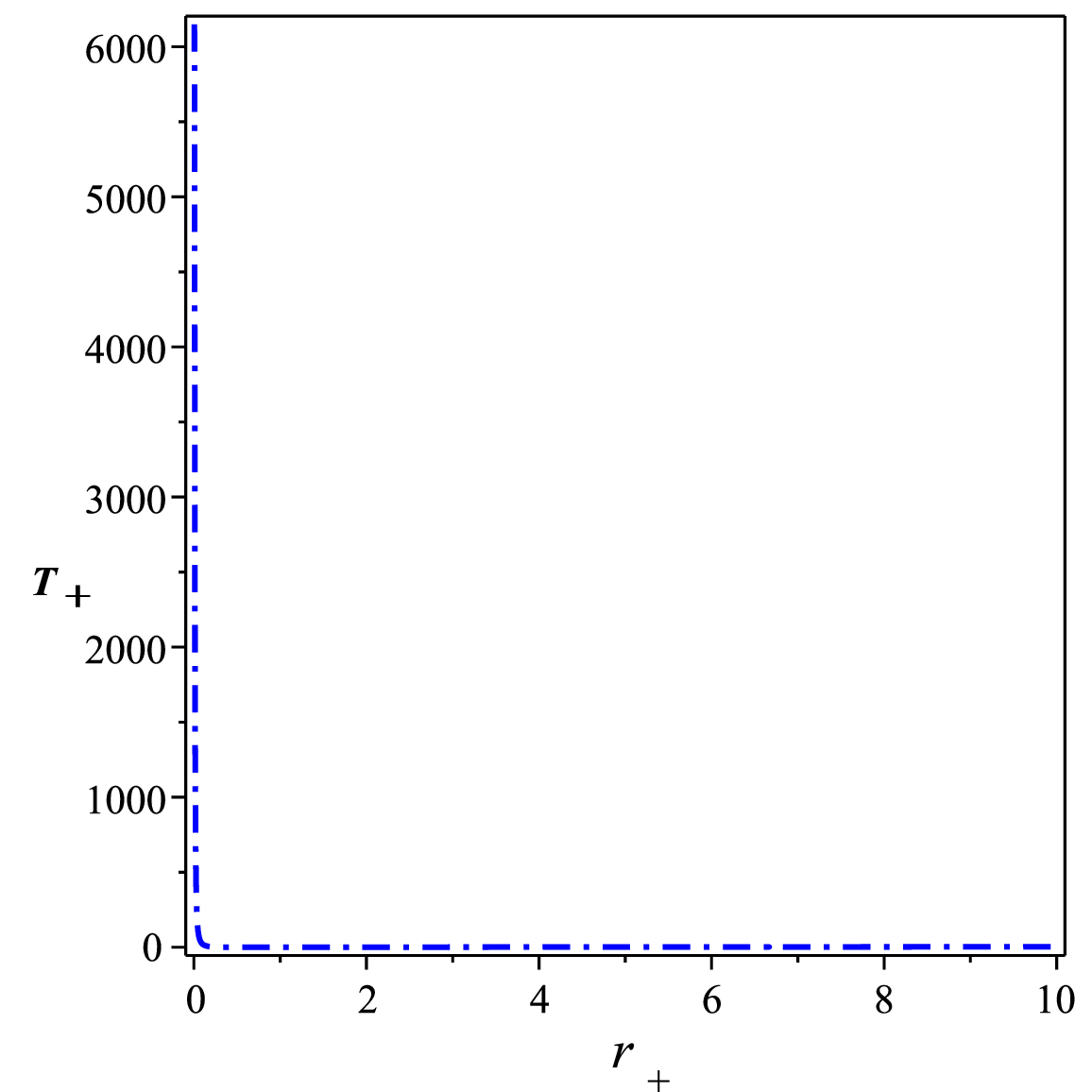}}
\subfigure[~The behavior of entropy given by Eq. (\ref{ent1})]{\label{fig:ent1}\includegraphics[scale=0.21]{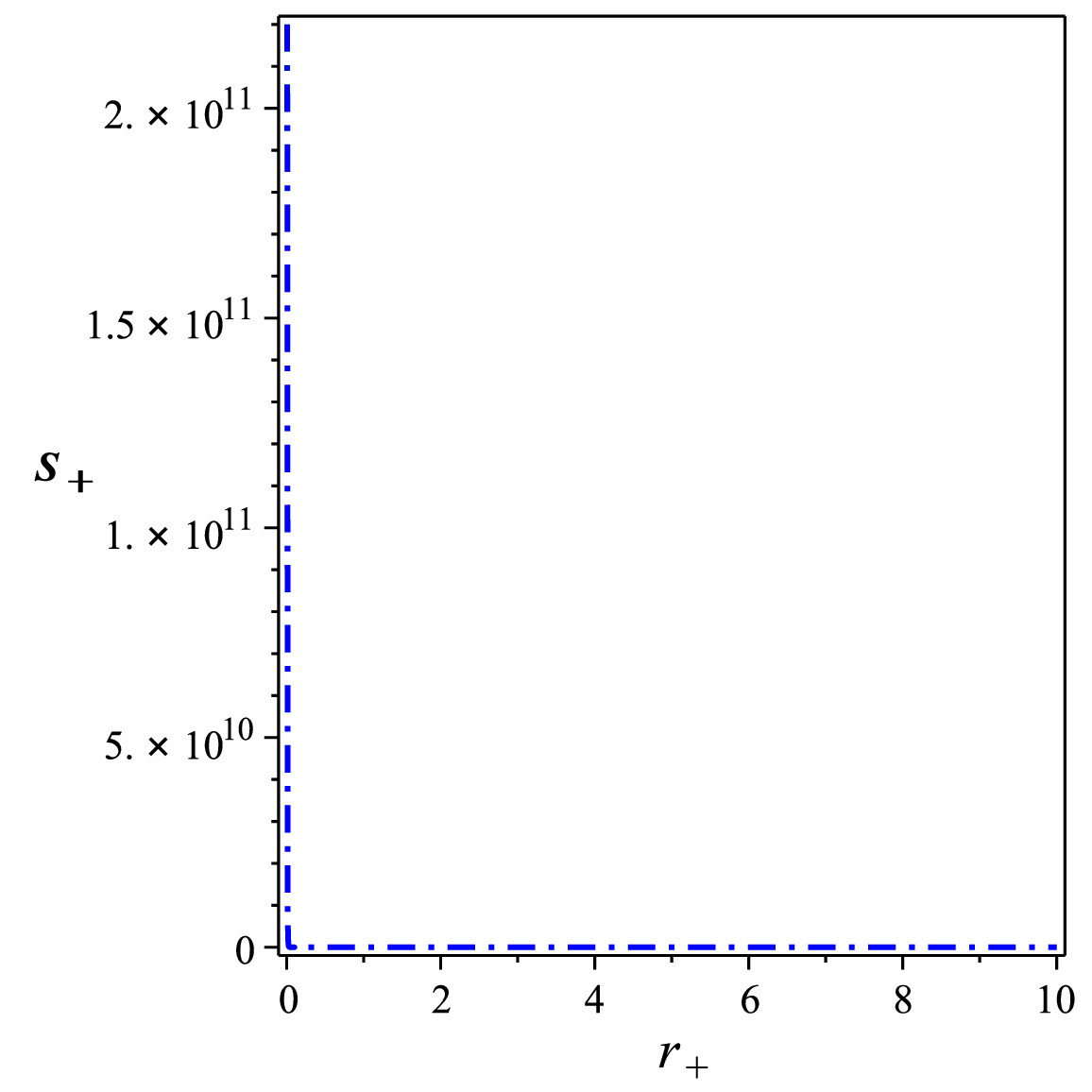}}
\subfigure[~The behavior of heat capacity given by Eq. (\ref{heat1})]{\label{fig:heat}\includegraphics[scale=0.21]{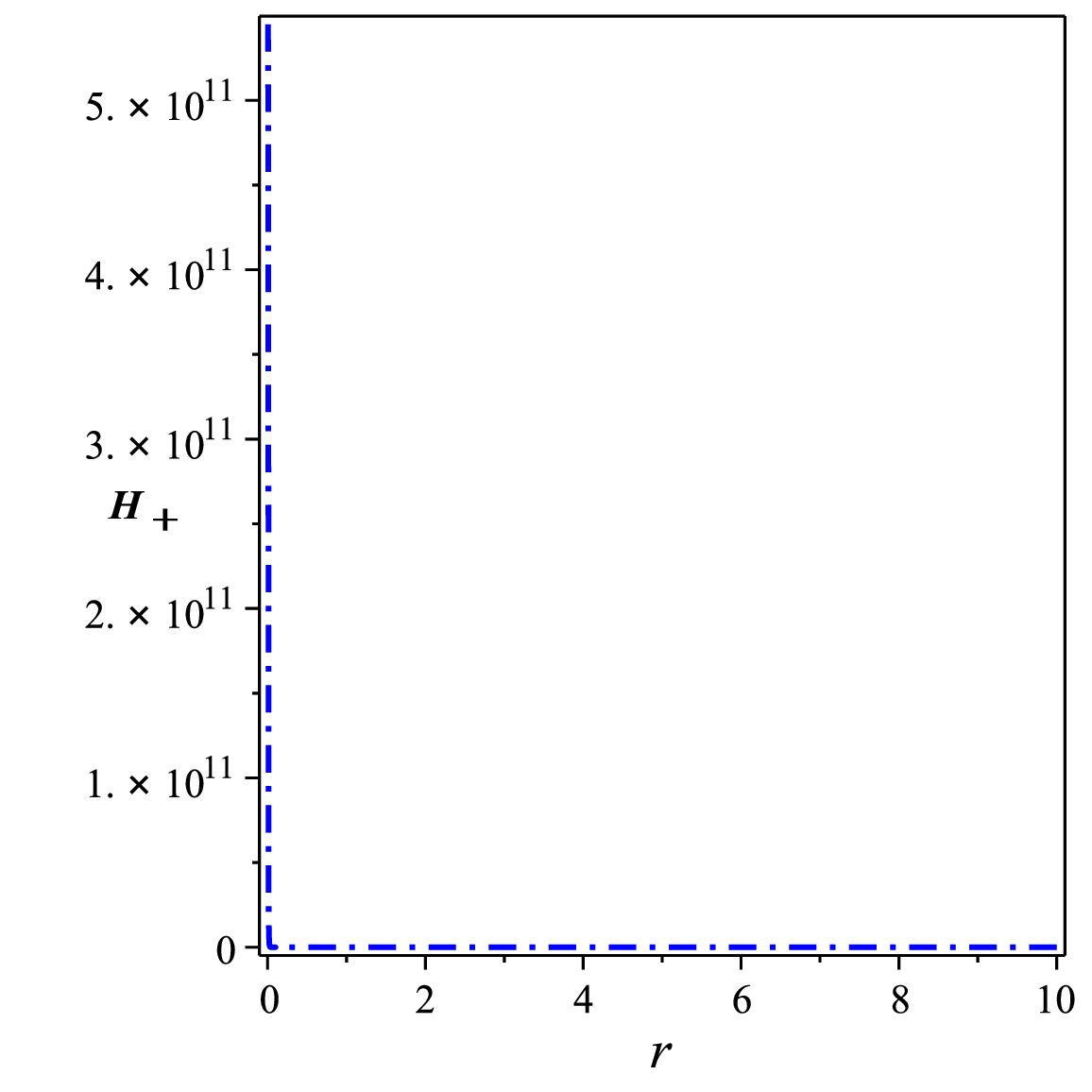}}
\subfigure[~The behavior of temperature given by Eq. (\ref{temp})]{\label{fig:temp}\includegraphics[scale=0.21]{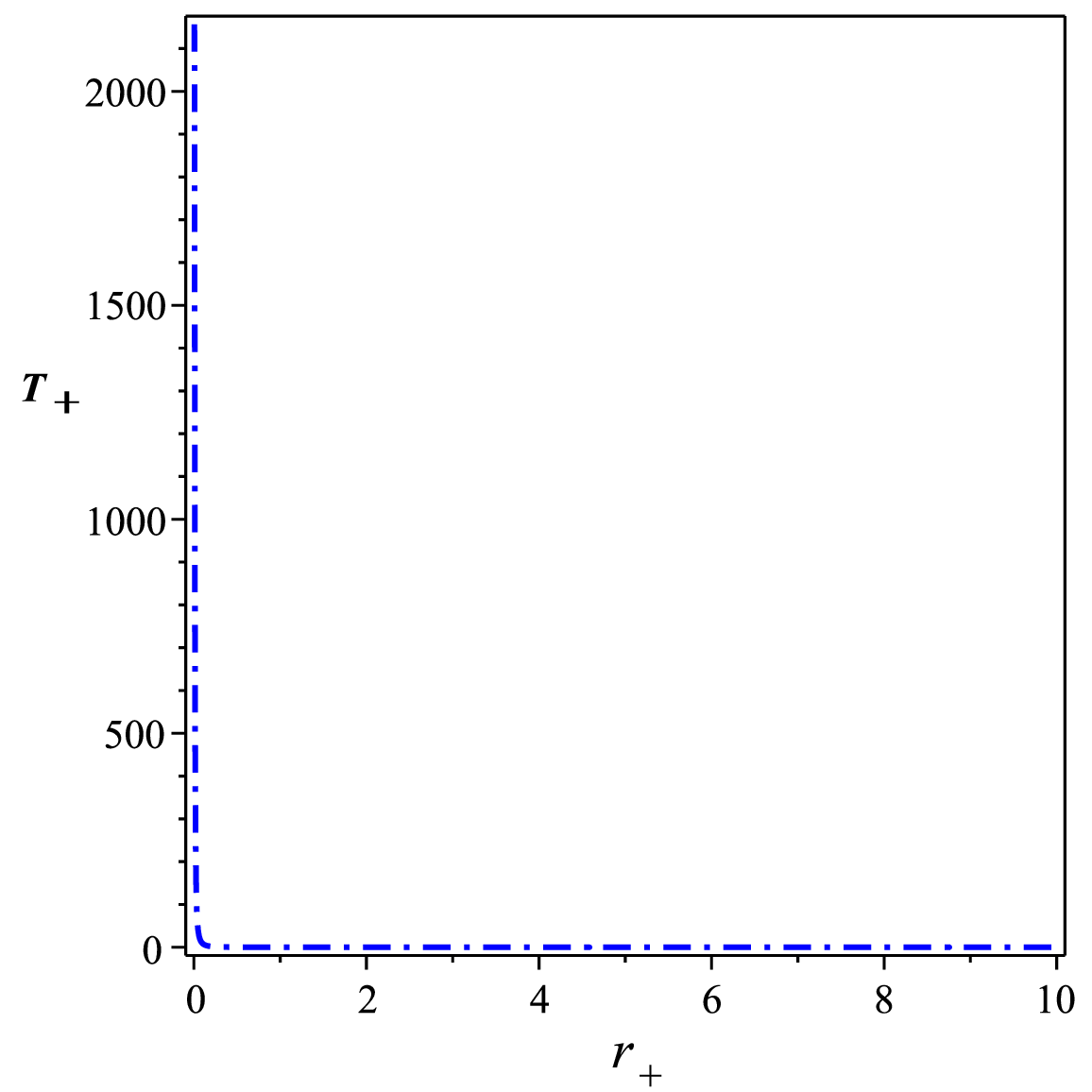}}
\subfigure[~The behavior of entropy for the case $\alpha=\frac{1}{12\Lambda}$]{\label{fig:ent11}\includegraphics[scale=0.21]{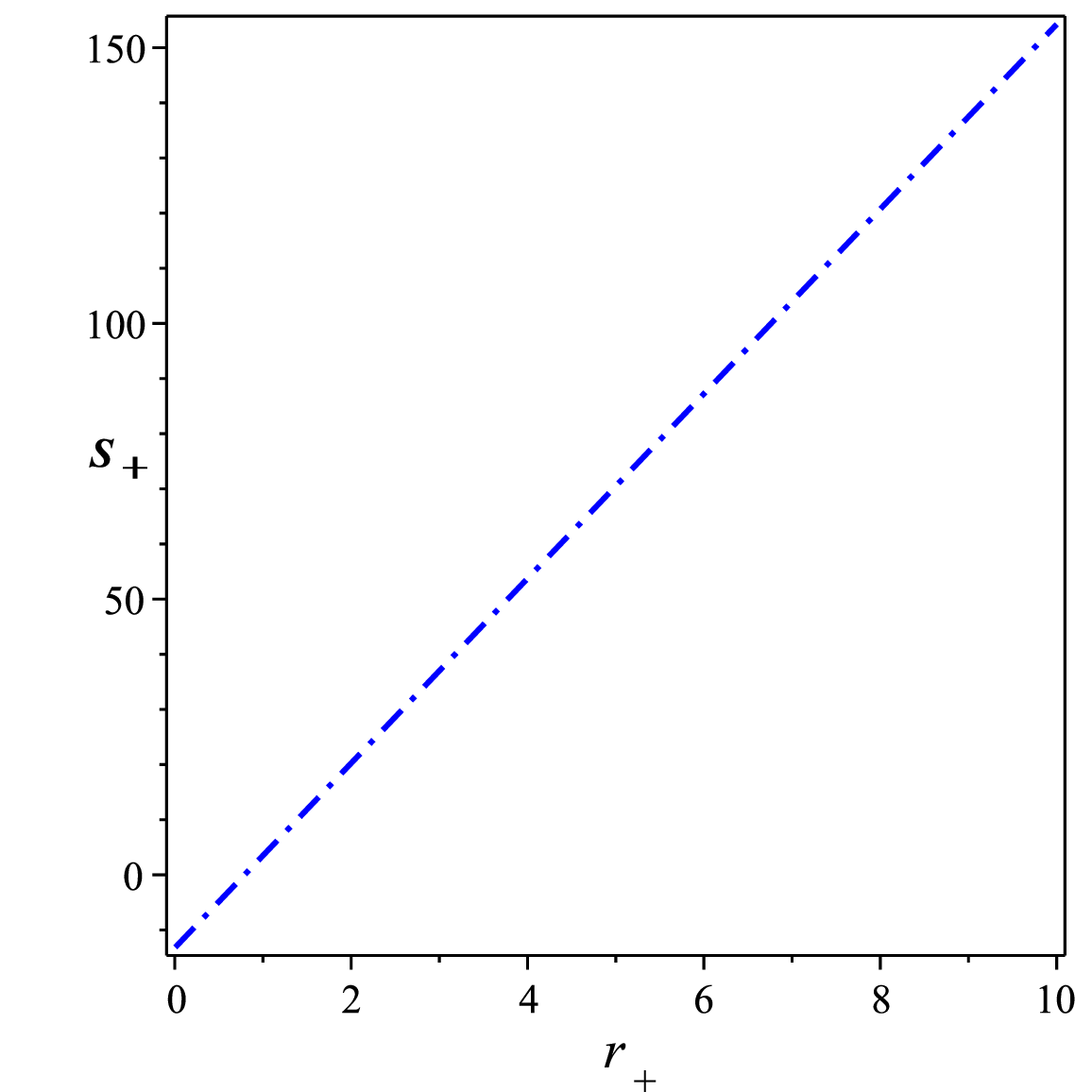}}
\subfigure[~The behavior of heat capacity for the case $\alpha=\frac{1}{12\Lambda}$]{\label{fig:heat11}\includegraphics[scale=0.21]{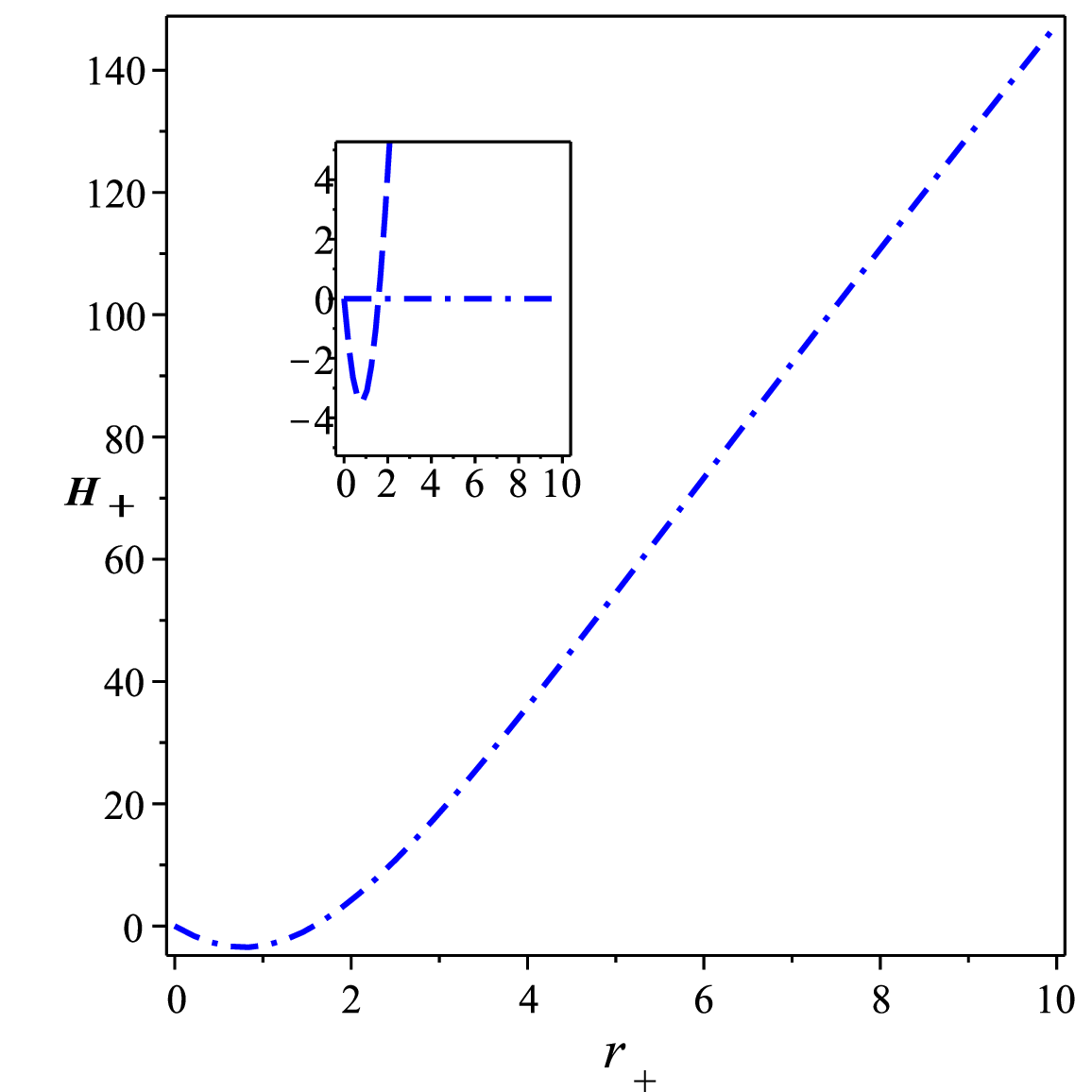}}
\caption{{~Thermodynamic properties of the black hole solutions.
Panels \subref{fig:temp1}--\subref{fig:heat} correspond to the generic quadratic $f(\mathbb{Q})$ branch with
$\alpha\neq1/(12\Lambda)$: \subref{fig:temp1} Hawking temperature given by Eq.~(\ref{kGR}),
\subref{fig:ent1} entropy given by Eq.~(\ref{ent1}), and (\ref{fig:heat}) heat capacity given by Eq.~(\ref{heat1}),
all plotted as functions of the horizon radius $r_{+}$.
Panels \subref{fig:temp}--\subref{fig:heat11} show the corresponding quantities for the special branch
$\alpha=1/(12\Lambda)$: \subref{fig:temp} Hawking temperature from Eq.~(\ref{temp}),
\subref{fig:ent11} entropy, and \subref{fig:heat11} heat capacity whose equations are given by (\ref{ent11}) and (\ref{heat11}).
The plots highlight the effects of quadratic non-metricity corrections on
thermodynamic stability, while the General Relativity (BTZ) limit is recovered
smoothly as $\alpha\to0$.
All curves are obtained for $\Lambda=-0.1$, $c_{2}=1$, and $\alpha=-0.1$.}}
\label{Fig:2}
\end{figure}
\begin{equation}
T(r_+) =\frac{\kappa}{2\pi} =\left. \frac{{ \sqrt{\nu_1(r)}}\mu'(r)}{4\pi}\right\vert _{r=r_{+}}\approx{\frac {36\,{r}^{3}{|\alpha|}^{3/2}\Lambda+{r}^{ 3}\sqrt {|\alpha|}+36c_2\,\sqrt {3}{\alpha}^{2}{r}^{2}\Lambda-3c_2\,\sqrt {3}\alpha\,{r}^{2}+36\,{|\alpha|}^{3/2}{c_2}^{2}r+24\,{c_2}^{3}\sqrt {3}{\alpha}^{2}}{144\pi \,{|\alpha|}^{5/2}{c_2}^{2}{r}^{2}}},  \label{kGR}
\end{equation}%
where $r_+$ is the radius of the event horizon of the AdS black hole. In Fig.~\ref{Fig:2} \subref{fig:temp1},  we show the behavior of the Hawking temperature given by Eq.  (\ref{kGR}). In this case,  the temperature is always positive. {
Equation~(\ref{kGR}) was obtained on the generic branch by solving the field equations with $\alpha\neq 0$, cf. Eq.~(\ref{gsol}), and
contains inverse powers of $|\alpha|$ (e.g.\ $|\alpha|^{-3/2}$ and $|\alpha|^{-5/2}$). Hence Eq.~(\ref{kGR}) is \emph{not} defined at
$\alpha=0$ and its naive substitution leads to a spurious divergence. Moreover, the horizon $r_+(\alpha)$ employed in
$T=\sqrt{\nu_1}\,\mu'(r_+)/4\pi$ is determined on that same branch by $\mu(r_+;\alpha,\Lambda,c_2,m)=0$ and does not
provide a uniform limit as $\alpha\to0$. Therefore the Einstein (BTZ) temperature must be computed by setting
$\alpha=0$ \emph{in the field equations} and then evaluating the surface gravity, yielding the BTZ result associated
with Eq.~(\ref{sol3}), rather than by taking $\alpha\to 0$ of Eq.~(\ref{kGR}).
}

The semi classical Bekenstein-Hawking entropy of the horizons is defined as\footnote{To determine the entropy of BTZ black holes, the area law can be applied as:
\begin{align}
&S=\frac{A}{4}, \quad \mbox{where $A$ is the horizon area and is defined by} \quad A=\left. \int_{0}^{2\pi }\sqrt{g_{\varphi \varphi }}d\varphi \right\vert
_{r=r_{+}}=\left. 2\pi r\right\vert _{r=r_{+}}=2\pi r_{+}, \quad  \mbox{where $g_{\varphi \varphi }=r^{2}$.} \label{AGR}
\end{align}
\begin{equation}
\mbox{So, the entropy of BTZ black holes is given by} \quad S=\frac{\pi r_{+}}{2}.  \label{EntGR}
\end{equation}
}:
\begin{align}\label{ent1}
&S(r_+)=\frac{{\cal A}}{4 } f_\mathbb{Q}(r_+)=\frac{1}2\pi  r_+f_\mathbb{Q}(r_+).
\end{align}
with ${\cal A}=2\pi r_+$  being the area of the event horizons\footnote{{
In $f(Q)$ gravity, the correct entropy expression follows from the N\"oetherWald approach and reads
$S = \frac{A}{4} f_Q(r_+)$, where $f_Q = \partial f / \partial Q = 1 + 2\alpha Q(r_+)$.
This reduces to the usual BekensteinHawking area law $S = A/4 = \pi r_+/2$ only in the Einstein limit
$\alpha \to 0$, where $f_Q \to 1$.
Therefore, Eqs.~(\ref{ent1}) and (\ref{AGR}) represent two consistent limits of the same framework:
Eq.~(\ref{ent1}) applies to the general quadratic $f(Q)$ theory, while Eq.~(\ref{AGR}) (the pure BTZ case)
emerges smoothly as $\alpha \to 0$.
This ensures internal consistency of the thermodynamics and justifies the use of both expressions.}}.  The asymptotic form of Eq.~\eqref{ent1} is listed in Appendix A. { Figure \ref{Fig:2}\subref{fig:ent1} demonstrates that the entropy calculated using Eq.~(\ref{ent1}) consistently remains positive, provided that the constraint  $1-12\alpha\Lambda\geq0$ is satisfied}.
Finally,   the heat capacity   is figured out as \cite{Zheng:2018fyn,Nashed:2019yto,Kim:2012cma}
\begin{align} \label{heat1}
&H(r_+)=T(r_+)\left(\frac{S'(r_+)}{T'(r_+)}\right)
\end{align}
where $S'(r_+)$ and $T'(r_+)$ are the derivative of entropy and Hawking temperature with respect to the outer horizon respectively.  We depict the behavior of the heat capacity given by Eq. (\ref{heat1})   in Fig.~\ref{Fig:2} \subref{fig:heat}. { It is important to note that, unlike the Hawking temperature in Eq.~(\ref{kGR}), both the entropy (\ref{ent1}) and the heat capacity (\ref{heat1}) remain finite in the limit $\alpha \to 0$.
This is because the entropy depends on $f_Q = 1 + 2\alpha Q$, which is analytic in $\alpha$ and smoothly reduces to the standard Bekenstein--Hawking area law $S = \pi r_+/2$.
Similarly, in the heat capacity expression, the powers of $|\alpha|$ appearing in $T'(r_+)$ and $S'(r_+)$ cancel, leading to a regular and finite result.
Therefore, the thermodynamic quantities derived from the area law and first law are well behaved at $\alpha=0$, reproducing the conventional BTZ black hole thermodynamics.}
\subsection{The case $\alpha=\frac{1}{12\Lambda}$}
In the case $\alpha=\frac{1}{12\Lambda}$, the Hawking temperature relation, through the definition of surface gravity ($\kappa$), is:
\begin{equation}
T(r_+) \approx -{\frac {-4\,{{\Lambda}}^{2}{r}^{3}-3\,{c_2}^{2}\Lambda r+{c_2}^{3}\sqrt {{|\Lambda|}}}{\pi\,{
r}^{2}}}
,  \label{temp}
\end{equation}%

In Fig.~\ref{Fig:2} \subref{fig:temp},  the behavior of the Hawking temperature given by Eq.  (\ref{temp}) is reported. Also in this case, the temperature is always positive.
 The behavior  of the entropy of the case $\alpha=\frac{1}{12\Lambda}$ is depicted in Fig.~\ref{Fig:2} \subref{fig:ent1}  showing that the  solution of the case $\alpha=\frac{1}{12\Lambda}$ has always positive entropy.
 We depict the behavior of the heat capacity given of the case $\alpha=\frac{1}{12\Lambda}$   in Fig.~\ref{Fig:2} \subref{fig:heat}.

{
Let $r_{+}$ be the largest root of the lapse,
\begin{equation}
\mu(r_{+};m,c_{2},\alpha,\Lambda)=0,
\label{eq:horizon}
\end{equation}
which defines the event horizon.  The integration constant $m$ entering $\mu(r)$ plays the role of the mass parameter; in $(2{+}1)$-D.  We may set
$M\equiv m$ up to a fixed normalization.
Varying the horizon condition \eqref{eq:horizon} at fixed $(\alpha,\Lambda)$ gives
\begin{equation}
0=d\mu|_{r_{+}}
=\underbrace{\mu'(r_{+})}_{4\pi T}\,dr_{+}
+\Bigl(\partial_{c_{2}}\mu\Bigr)_{+} dc_{2}
+\Bigl(\partial_{m}\mu\Bigr)_{+} dm.
\label{eq:varmu}
\end{equation}
 Hence
\begin{equation}
dm=\mu'(r_{+})\,dr_{+}+\Bigl(\partial_{c_{2}}\mu\Bigr)_{+} dc_{2}
=4\pi T\,dr_{+}+\Bigl(\partial_{c_{2}}\mu\Bigr)_{+} dc_{2}, \quad \mbox{where s $(\partial_{m}\mu)_{+}=-1$}.
\label{eq:dm}
\end{equation}
Using Eq.~\eqref{ent11} we get,
\begin{equation}
dS=\frac{\pi}{2}\Bigl[\,f_{Q}(r_{+})\,dr_{+}+r_{+}\,df_{Q}(r_{+})\Bigr]
=\frac{\pi}{2}\Bigl[\,f_{Q}\,dr_{+}+r_{+}\alpha\,dQ(r_{+})\Bigr],
\label{eq:dS}
\end{equation}
with
\(
dQ(r_{+})
=Q'_{+}\,dr_{+}
+\bigl(\partial_{c_{2}}Q\bigr)_{+}\,dc_{2},
\)
where the prime denotes $d/dr$.
Using \eqref{5} one can show that the combination appearing in $TdS$ reproduces the
$dr_{+}$ term in \eqref{eq:dm}. In particular, inserting \eqref{1} and \eqref{5},
\begin{align}
T\,dS
&=\frac{\mu'(r_{+})}{4\pi}\cdot \frac{\pi}{2}
\Bigl[f_{Q}\,dr_{+}+r_{+}\alpha\,\bigl(Q'_{+}\,dr_{+}+(\partial_{c_{2}}Q)_{+}\,dc_{2}\bigr)\Bigr]\nonumber\\
&=\frac{\mu'(r_{+})}{8}\Bigl[\underbrace{f_{Q}+r_{+}\alpha Q'_{+}}_{\displaystyle \mathcal{K}_{+}}\Bigr]dr_{+}
+\frac{\mu'(r_{+})}{8}\,r_{+}\alpha\,(\partial_{c_{2}}Q)_{+}\,dc_{2}.
\label{eq:TdSraw}
\end{align}
The horizon-field equations \eqref{1}--\eqref{5} imply the identity
\begin{equation}
\mathcal{K}_{+}=2,
\qquad\text{so that}\qquad
T\,dS=\mu'(r_{+})\,\frac{dr_{+}}{4}
+\frac{\mu'(r_{+})}{8}\,r_{+}\alpha\,(\partial_{c_{2}}Q)_{+}\,dc_{2}.
\label{eq:Kdentity}
\end{equation}
Combining \eqref{eq:dm} and \eqref{eq:Kdentity} we obtain
\begin{equation}
dm = T\,dS \;+\;\underbrace{\Bigl[\;(\partial_{c_{2}}\mu)_{+}-\frac{\mu'(r_{+})}{8}\,r_{+}\alpha\,(\partial_{c_{2}}Q)_{+}\;\Bigr]}_{\displaystyle \Psi}\,dc_{2}.
\label{eq:firstlaw}
\end{equation}
Identifying $M\equiv m$ and the intensive conjugate $\Psi$ to the charge-like hair $c_{2}$, \eqref{eq:firstlaw} is the
\emph{first law}
\begin{equation}
dM = T\,dS + \Psi\,dc_{2},
\qquad (\alpha,\Lambda\ \text{fixed}).
\end{equation}

\paragraph*{Checks on the two branches.}
\begin{itemize}
\item For the generic branch $\alpha\neq 1/(12\Lambda)$, $T$ is given in Eq.\,(42) and $S$ in Eq.\,(45); inserting those
expressions into \eqref{eq:firstlaw} reproduces $dm$ obtained from the variation of $\mu(r_{+})=0$.
\item For the special branch $\alpha=1/(12\Lambda)$, one uses Eq.\,(47) for $T$ and Eq.\,(50) for $S$, and the same
\end{itemize}
We should note the following:\\
(i) In the Einstein (BTZ) limit ($\alpha\to 0$, $\nu_{1}\to 1$) one has $f_{Q}\to 1$, $Q\to -2\Lambda$, so $S\to \tfrac{1}{2}\pi r_{+}$
and $\Psi\to (\partial_{c_{2}}\mu)_{+}$, reducing the first law to the familiar BTZ relation with the electric contribution
encoded by $c_{2}$. (ii) If one chooses to keep $\Lambda$ or $\alpha$ variable, their conjugates appear in the first law
as $+\;V\,dP$ with $P=-\Lambda/(8\pi)$ (extended thermodynamics) and $+\;\Theta\,d\alpha$, obtainable by varying
\eqref{eq:horizon} at fixed $(r_{+},c_{2})$.
}

\section{ Comparing  modified gravity   and General Relativity by the geodesic equations}\label{sec:44}

In this section, we study the geodesic motions of test particles in the BTZ  metric (\ref{met222}) and compare the results with those obtained in General Relativity.

The geodesics equations are given by
\begin{equation}
    \ddot{x}^{\lambda}+\Gamma^{\lambda}_{\mu\nu}\,\dot{x}^{\mu}\,\dot{x}^{\nu}=0,\label{A1}
\end{equation}
where dot represents derivative w. r. t. the affine parameter $\tau$.

The metric tensor $g_{\mu\nu}$ for the space-time (\ref{met222}) depends on the coordinate $r$ only, and it is independent of the coordinate $t, \phi$. Therefore,  two Killing vectors $\partial_{t}=\frac{\partial}{\partial t}$ and $\partial_{\phi}=\frac{\partial}{\partial \phi}$ exist. The corresponding constants of motion with respect to the parameter $\tau$ can be derived using the relation $k=g_{\mu\nu}\,\xi^{\mu}\,\frac{dx^{\nu}}{d\tau}$\footnote{ {Here $k$ represents the  conserved quantity associated with the symmetry generated by the Killing vector field $\xi^\mu$.}}. These constants $(\mathrm{E},\mathrm{L})$ are given by
\begin{eqnarray}
    -\mathrm{E}=g_{tt}\,\dot{t}, \qquad \qquad
    \mathrm{L}=g_{\phi\phi}\,\dot{\phi}.\label{A2}
\end{eqnarray}
From those,  we get
\begin{eqnarray}
\dot{t}=-\frac{\mathrm{E}}{g_{tt}},
\qquad \qquad
    \dot{\phi}=\frac{\mathrm{L}}{g_{\phi\phi}}=\frac{\mathrm{L}}{r^2},\label{A3}
\end{eqnarray}
where $g_{tt}$ and $g_{\phi,\phi}$ are defined in Eq.~\eqref{met222}.
It is worth mentioning that the denominator of $\dot{\phi}$ vanish at the  horizon $r=r_+$, while, for naked singularities, both  $\dot{t}$ and  $\dot{\phi}$ are positive defined. Consequently, the geodesics around  naked singularities are drastically different from those in the black hole case.

The Lagrangian of the system is defined by
\begin{equation}
    \mathcal{L}=\frac{1}{2}\,g_{\mu\nu}\,\frac{dx^{\mu}}{d\tau}\,\frac{dx^{\nu}}{d\tau}.\label{A5}
\end{equation}
Using the metric tensor (\ref{met222}), we find
\begin{eqnarray}
    g_{tt}\,\dot{t}^2+g_{rr}\,\dot{r}^2+g_{\phi\phi}\,\dot{\phi}^2=\epsilon,\label{A6}
\end{eqnarray}
where $\epsilon=0$ for null geodesics and $1$ for time-like geodesics.
Substituting metric tensor components and after simplification, we can write
\begin{eqnarray}
    \dot{r}^2=\frac{\epsilon}{g_{rr}}-\frac{\mathrm{E}^2}{g_{tt}g_{rr}}-\frac{L^2}{r^2g_{rr}}.\label{A7}
\end{eqnarray}
To simplify the calculations  we assume that $g_{tt}g_{rr}\approx -1$ which comes from the fact that $\nu_1(r)=1$ which is satisfied when the constant $c_3=-4\Lambda$ and $c_2=0$. When $g_{tt}g_{rr}\approx -1$  Eq. \eqref{A7} can be rewritten as:
\begin{eqnarray}
    \dot{r}^2=\frac{\epsilon}{g_{rr}}+\mathrm{E}^2-\frac{L^2}{r^2g_{rr}}.\label{A77}
\end{eqnarray}
The effective potential of the system is given by
\begin{eqnarray}
    V_{eff} (r)&=&-\frac{\epsilon}{g_{rr}}+\frac{L^2}{r^2g_{rr}}, \label{A8}
\end{eqnarray}
where $\Lambda$ is connected through $\alpha$ by $\alpha\Lambda=\frac{1}{12}$. We see that the potential of the system is influenced by the coupling constant $\alpha$ involved in the cosmological constant. This dimensional constant alter the effective potential of the system compared to the result in General Relativity.
\begin{figure}
\centering
\subfigure[~The behavior of the potential given by Eq. \eqref{A8} when $\epsilon=0$]{\label{fig:VL}\includegraphics[scale=0.24]{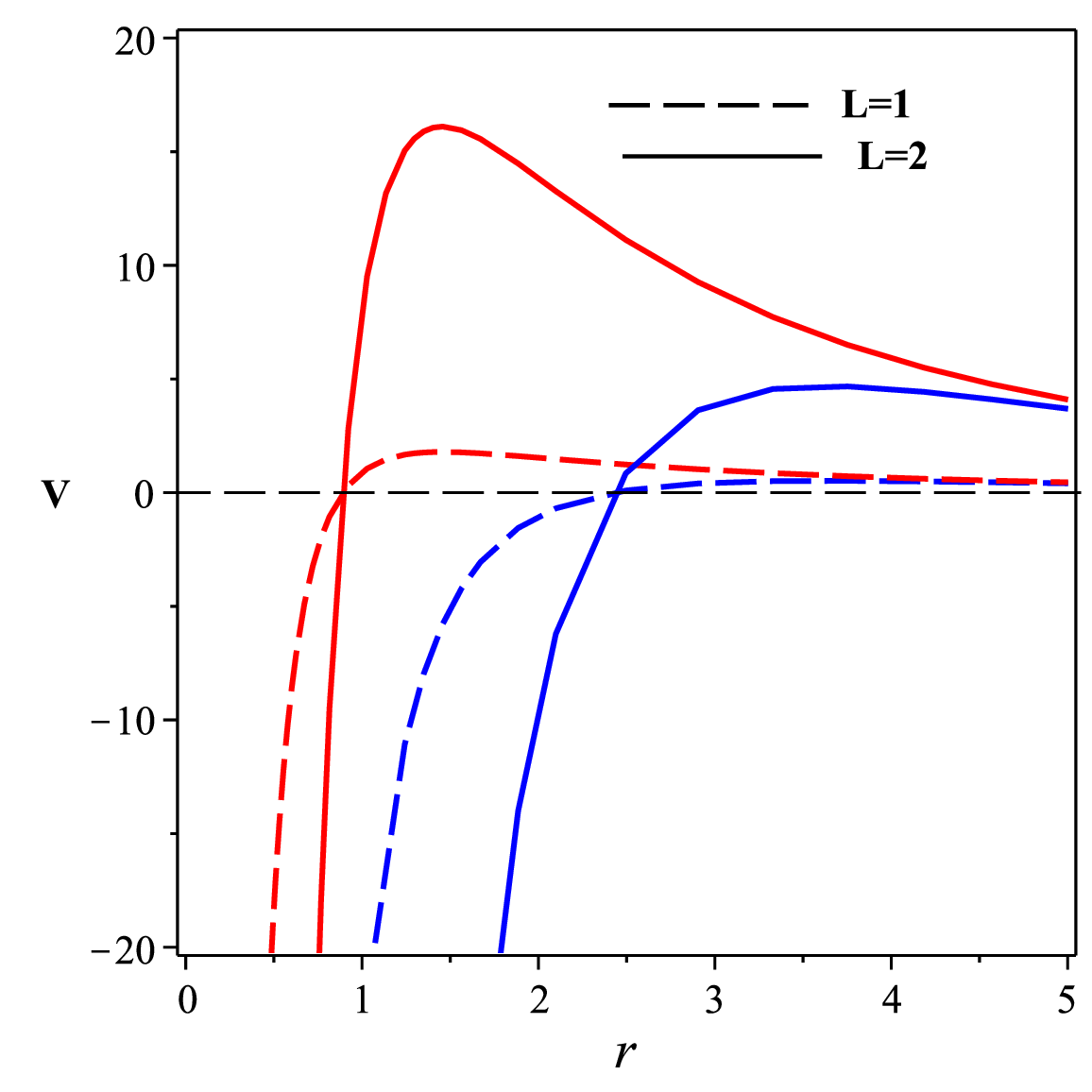}}
\subfigure[~The behavior of the potential given by  Eq. \eqref{A8} when $L=1$]{\label{fig:VE}\includegraphics[scale=0.24]{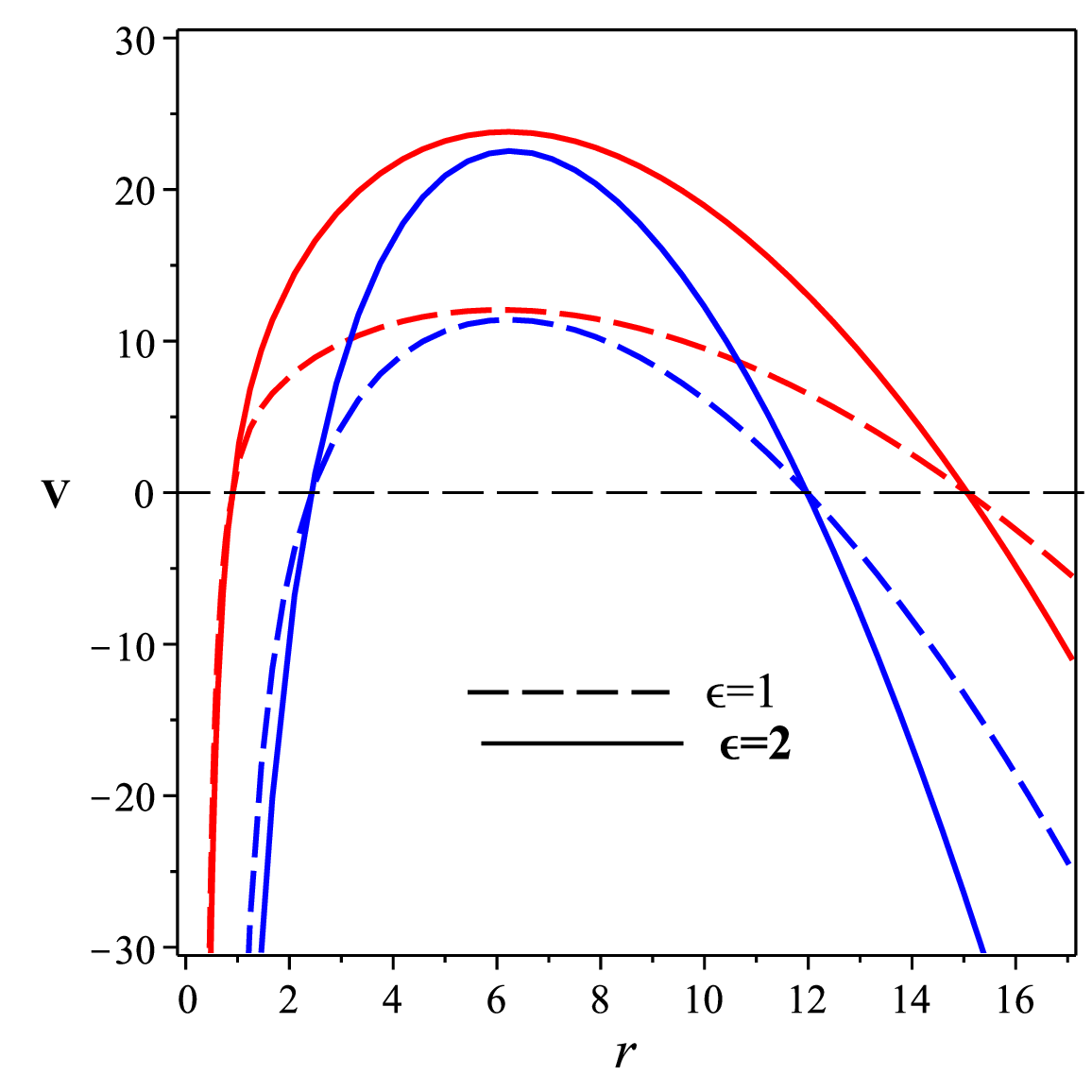}}
\subfigure[~The behavior of the potential given by Eq. \eqref{A8} ]{\label{fig:ddv}\includegraphics[scale=0.24]{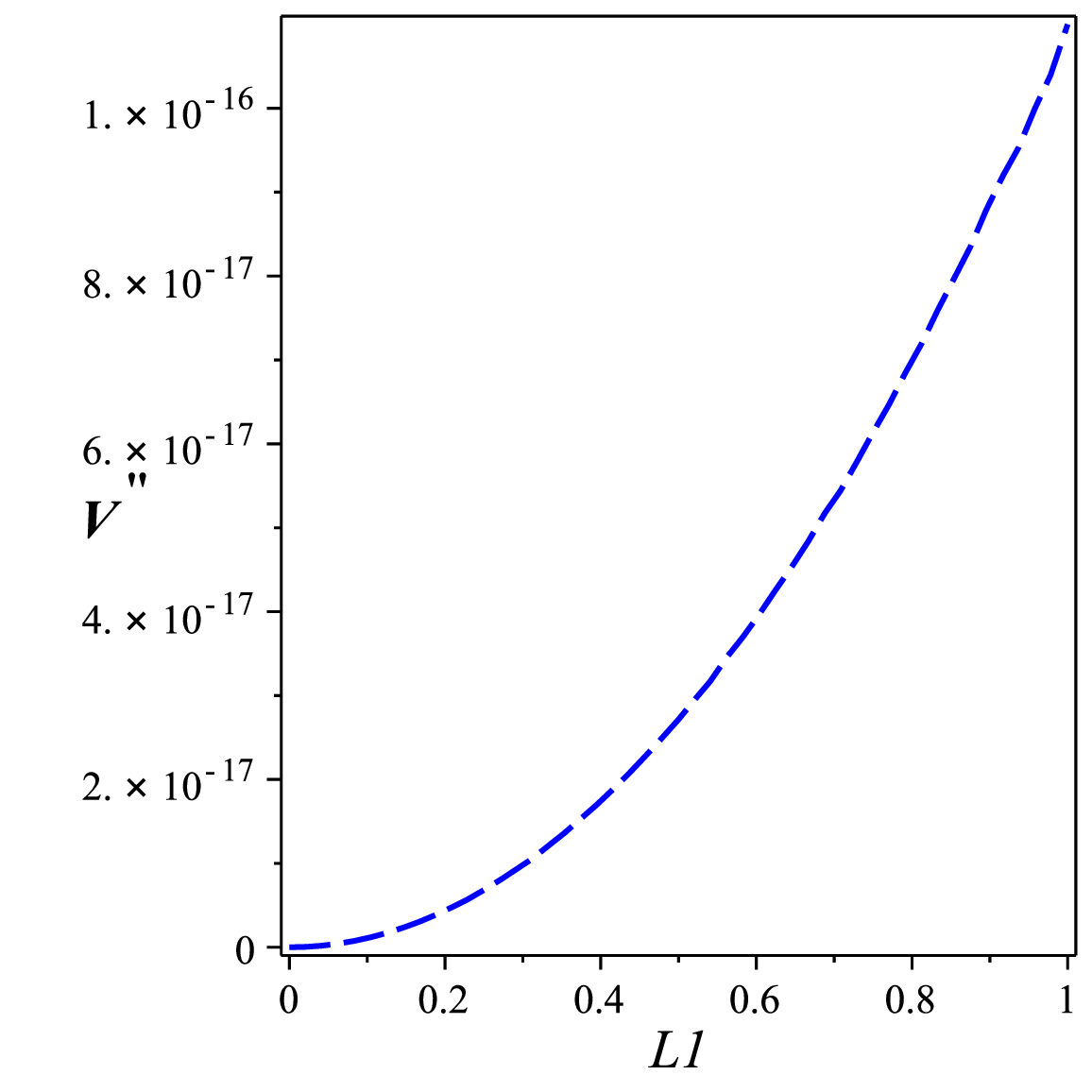}}
\caption{{Effective potential governing geodesic motion.
Panel \subref{fig:VL} shows the effective potential $V_{\mathrm{eff}}$ for null geodesics ($\epsilon=0$)
for different values of the angular momentum $L$.
The red curves correspond to the General Relativity (BTZ) case $\alpha=0$,
while the blue curves represent the quadratic $f(\mathbb{Q})$ gravity case with $\alpha\neq0$.
Panel \subref{fig:VE} displays the effective potential for time-like geodesics ($\epsilon\neq0$),
showing analogous deviations induced by non-metricity corrections.
Panel \subref{fig:ddv} shows the second derivative of the effective potential, indicating the
stability properties of circular photon orbits.
The comparison clearly demonstrates how quadratic $f(\mathbb{Q})$ corrections
modify particle dynamics relative to the BTZ geometry.}}

\label{Fig:3}
\end{figure}

 Fig. \ref{Fig:3} illustrates the behavior of the effective potential for null geodesics. In panel \subref{fig:VL}  of Fig. \ref{Fig:3}, the dotted red line, representing $\mathrm{L}=1$, corresponds to the General Relativity case, $\alpha=0$. The dotted blue line is for  $\alpha\neq 0$, where the constants are set to $c_2=2$, $\Lambda=-0.01$ and $c_3=0.4$. In contrast, the solid red line, for $\mathrm{L}=2$, represents $\alpha=0$,  and the blue curve represents $\alpha\neq 0$.

 Fig.~\ref{Fig:3},  panel \subref{fig:VE} ,  illustrates the behavior of the effective potential for time-like geodesics in both $\alpha=0$ and $\alpha\neq 0$. The explanation of this is analogue to the previous Fig.~\ref{Fig:3} panel \subref{fig:VL} .

Finally, in Fig.~\ref{Fig:3} panel \subref{fig:ddv},  illustrates the behavior of the second derivative of effective potential for null geodesics  which shows the stability of the photon orbits.
\section{Multi horizons}\label{multi}
    We now discuss how to generate black holes with multi horizons. To do this, we will use the solution given by Eq~\eqref{gsol} where $\alpha\neq \frac{1}{12\Lambda}$. The asymptotic behaviour of this solution is given by
   \begin{align}\label{ggsol}
   &\mu=-\frac{1}2{\frac { \left( \alpha\Lambda+\frac{1}{36} \right) {r}^{2}}{{ \alpha}^{2}{c_2}^{2}c_3}}+\frac{1}{12}{\frac { \left( 12\,{ \alpha}^{2}{c_2}^{2}\Lambda-\alpha\,{c_2}^{2} \right) \sqrt {-3\,\alpha\,{c_2}^{2}}r}{{\alpha}^{3}{c_2}^{4}c_3}}-\frac{1}2\,{\frac {2\,{c_2}^{2}mc_3\,{ \alpha}^{2}+2\,\alpha\,{c_2}^{2}\ln  \left( \frac{r}{r_0}\right) }{{\alpha }^{2}{c_2}^{2}c_3}}-\frac{2}3\,{\frac {\sqrt {-3\,\alpha\,{c_2}^{2}}}{c_3\,\alpha\,r}}\,,\nonumber\\
   &\nu_1={\frac {1}{96{\alpha}^{2}{c_2}^{6}}} \left[ -162\alpha{c_2}^{2}  \ln \left( \frac{r}{r_0}\right) -16{c_2}^{2} \left\{ 9 \left(c_3c_4 -2\Lambda \right) {\alpha}^{2}+ 3\alpha+{\frac {81}{8}}{\alpha}^{3}{c_4}^{2}{c_3}^{2} \right\} \ln  \left( \frac{r}{r_0}\right)-162 {c_2}^{2}\alpha \left( c_4c_3\alpha+\frac{8}9 \right)   \ln  \left( \frac{r}{r_0}\right)\right.\nonumber\\
    &\left.-16\alpha {c_2}^{2}\left( {\frac {27}{8}}{\alpha}^{3}{c_4}^{3}{ c_3}^{3}+\frac{9}2c_3c_4 \left( c_3c_4 -4\Lambda \right) {\alpha}^{2}+3 \left( c_3c_4-{\frac {19}{4}}\Lambda \right) \alpha+{\frac {15}{16}} \right)  \right] { r}^{2}+{\frac {1}{96{\alpha}^{2}{c_2}^{6}}} \left[ 36 \alpha\left\{ 2\ln \left( \frac{r}{r_0}\right)[1+ c_4c_3\alpha]\right.\right.\nonumber\\
     &\left.\left. + {c_4}^{2}{c_3}^{2}{\alpha}^{2}-\frac{10}3\alpha\Lambda+{\frac {13}{18}} \right\}{c_2}^{2}  \sqrt {-3\alpha{c_2}^{2}} -16 \left( -3\alpha{c_2}^{2} \right) ^{3/2}\ln  \left( \frac{r}{r_0}\right) -16\alpha \left( -3\alpha{c_2}^{2} \right) ^{3 /2}c_4c_3 \right] r\nonumber\\
     &+{ \frac {1}{12}}{\frac {-\alpha \left(-9{c_2}^{4}c_3c_4{\alpha}^{2}-12\alpha{c_2}^{4} \right) +9{\alpha}^{2}{c_2}^{4}\ln  \left( \frac{r}{r_0}\right) }{{ \alpha}^{2}{c_2}^{6}}}-\frac{1}2{\frac {\sqrt {-3\alpha{c_2}^{2}}}{{c_2}^{2}r}}\,.
   \end{align}
Eq. \eqref{ggsol} shows that the dimensional parameter should have a negative value. The behavior of  Eq.~\eqref{ggsol} is shown in  Fig.~\ref{Fig:4}.
\begin{figure}
\centering
\subfigure[~The behavior of the metric given by Eq. \eqref{ggsol} for the ansatz  $\mu$]{\label{fig:metm}\includegraphics[scale=0.21]{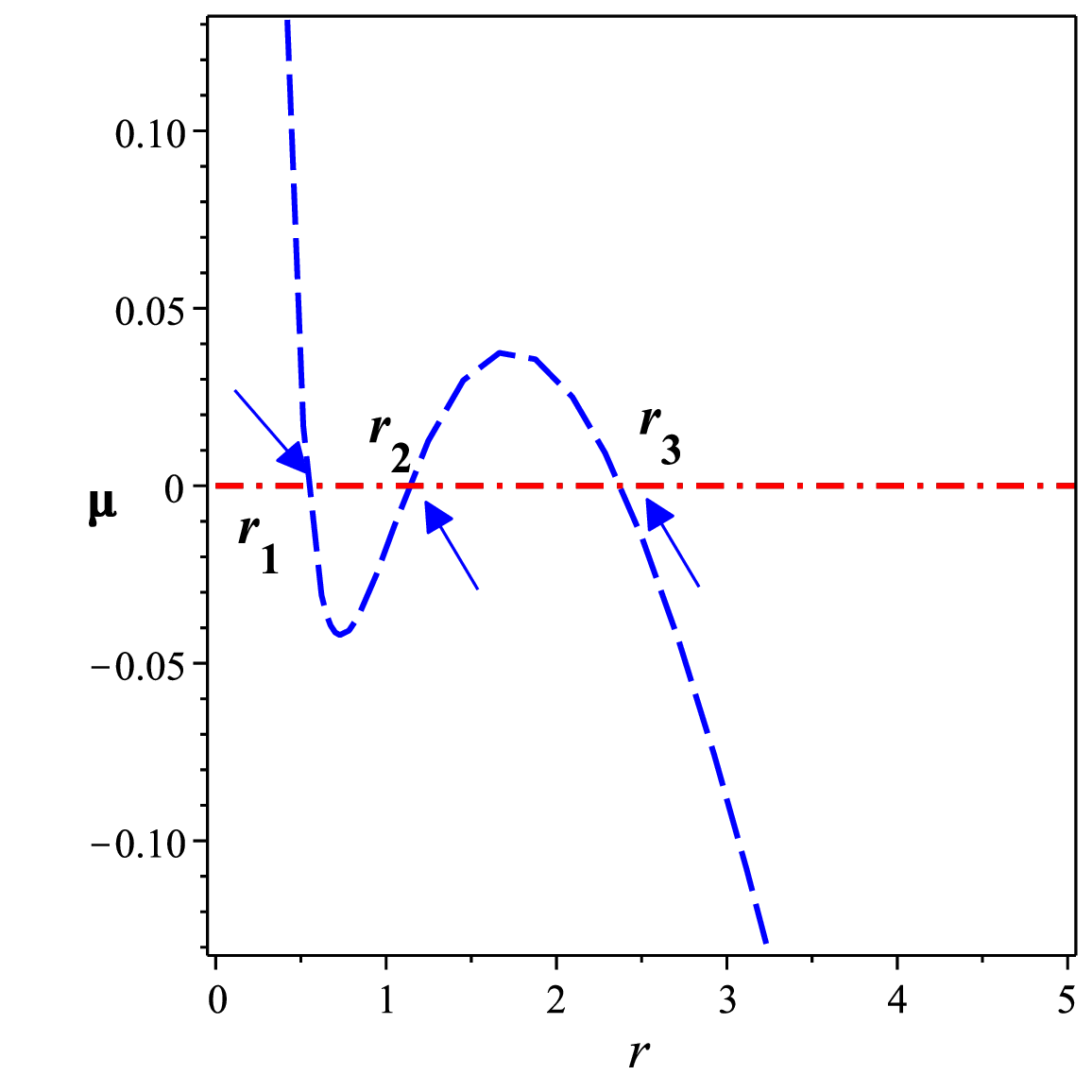}}
\subfigure[~The behavior of the metric given by  Eq. \eqref{ggsol} for the ansatz $\nu$]{\label{fig:metn}\includegraphics[scale=0.21]{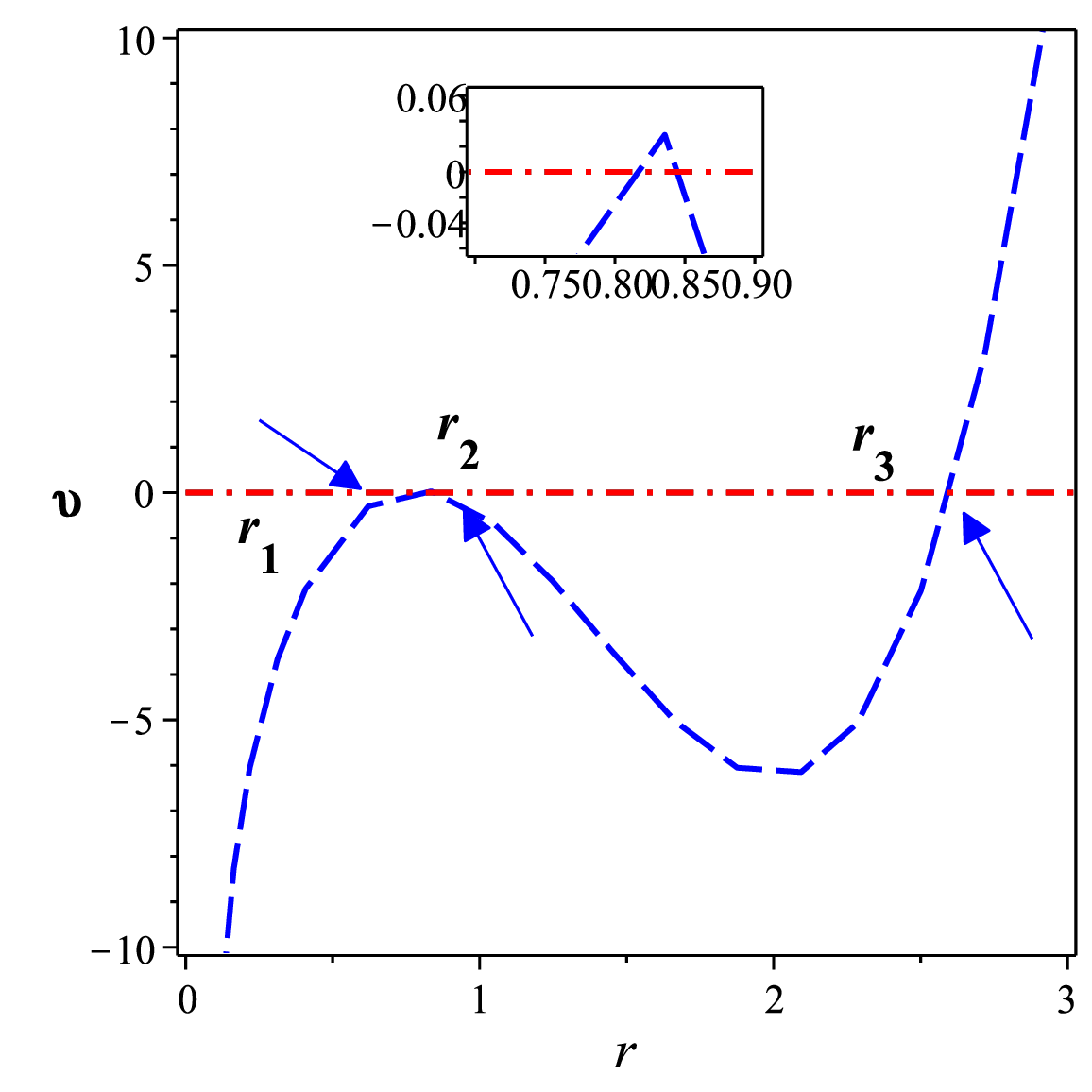}}
\subfigure[~The behavior of the Hawking temperature associated with the solution given by  Eq. \eqref{ggsol}]{\label{fig:tempm}\includegraphics[scale=0.21]{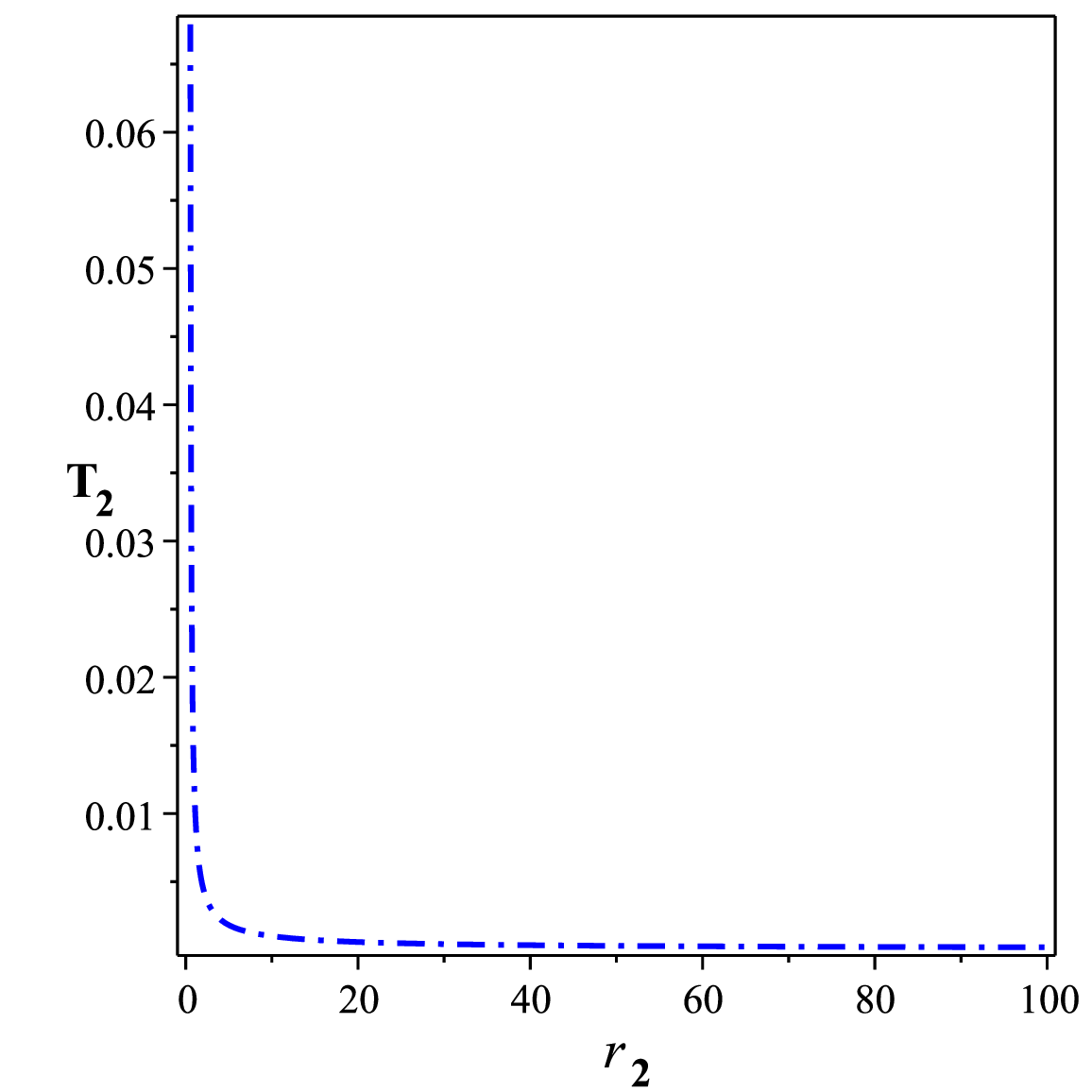}}
\subfigure[~The behavior of the entropy associated with the solution given by  Eq. \eqref{ggsol}]{\label{fig:entm}\includegraphics[scale=0.21]{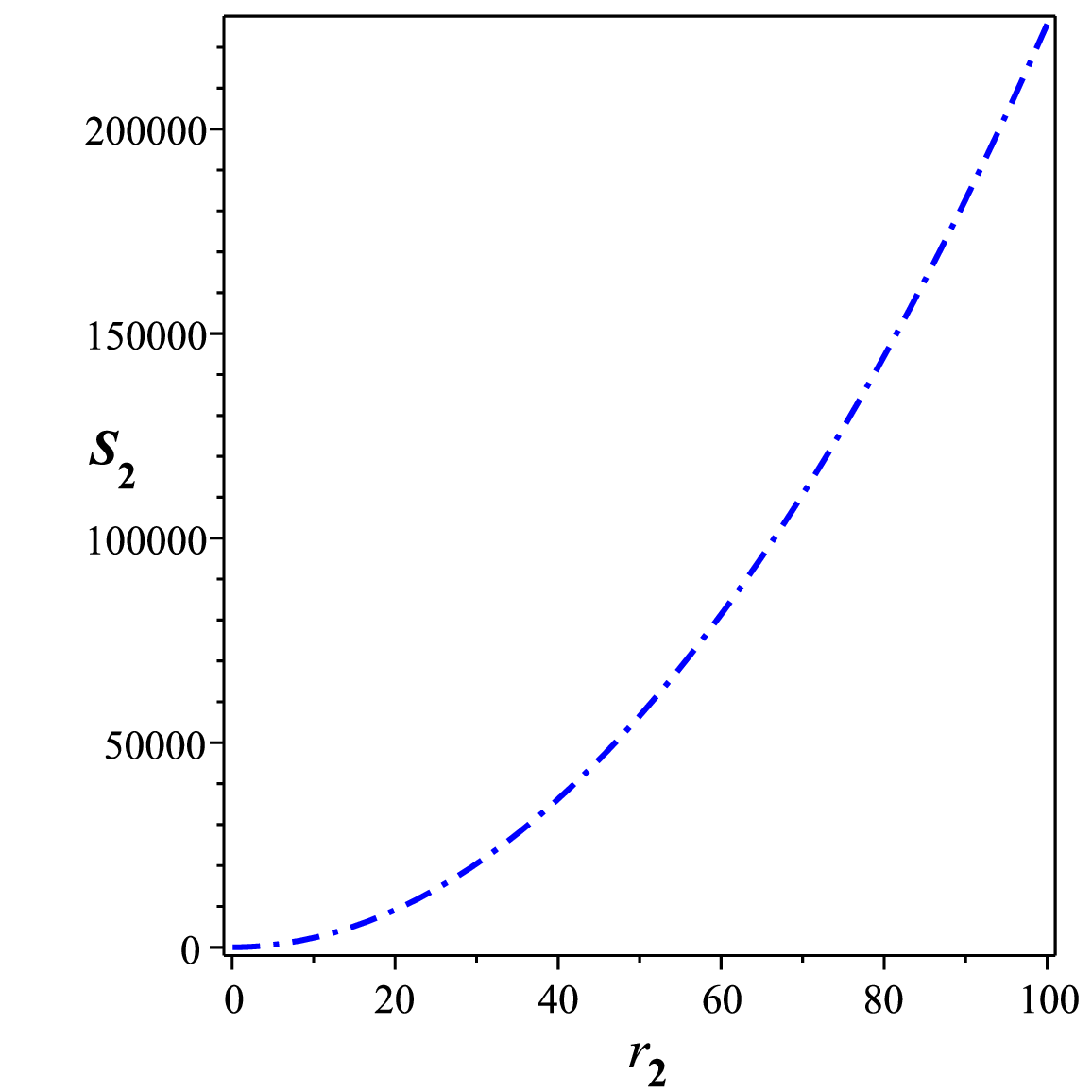}}
\subfigure[~The behavior of the heat capacity associated with the solution given by  Eq. \eqref{ggsol}]{\label{fig:heatm}\includegraphics[scale=0.21]{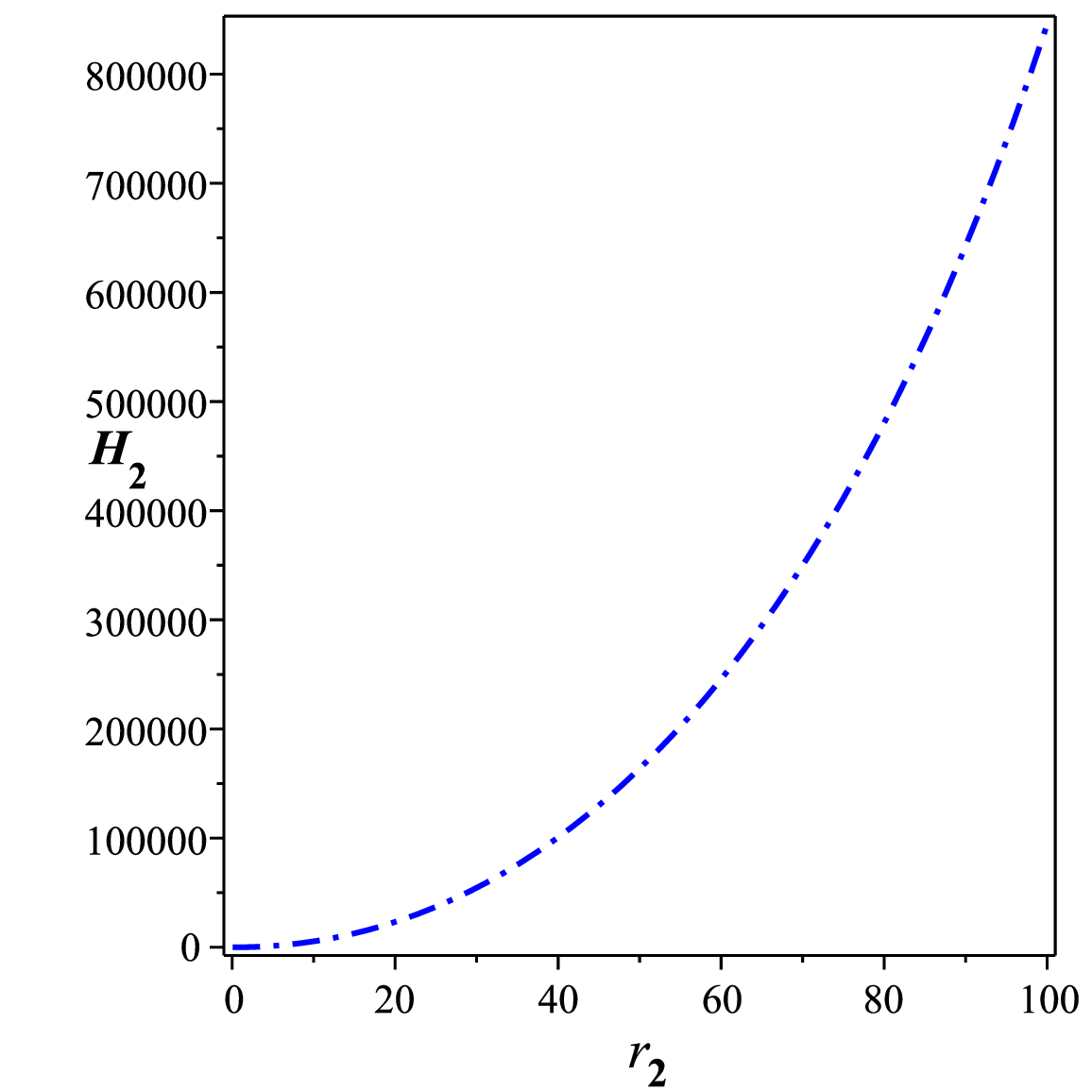}}
\subfigure[~The behavior of the second derivative of the potential associated with the solution given by  Eq. \eqref{ggsol}]{\label{fig:potm}\includegraphics[scale=0.21]{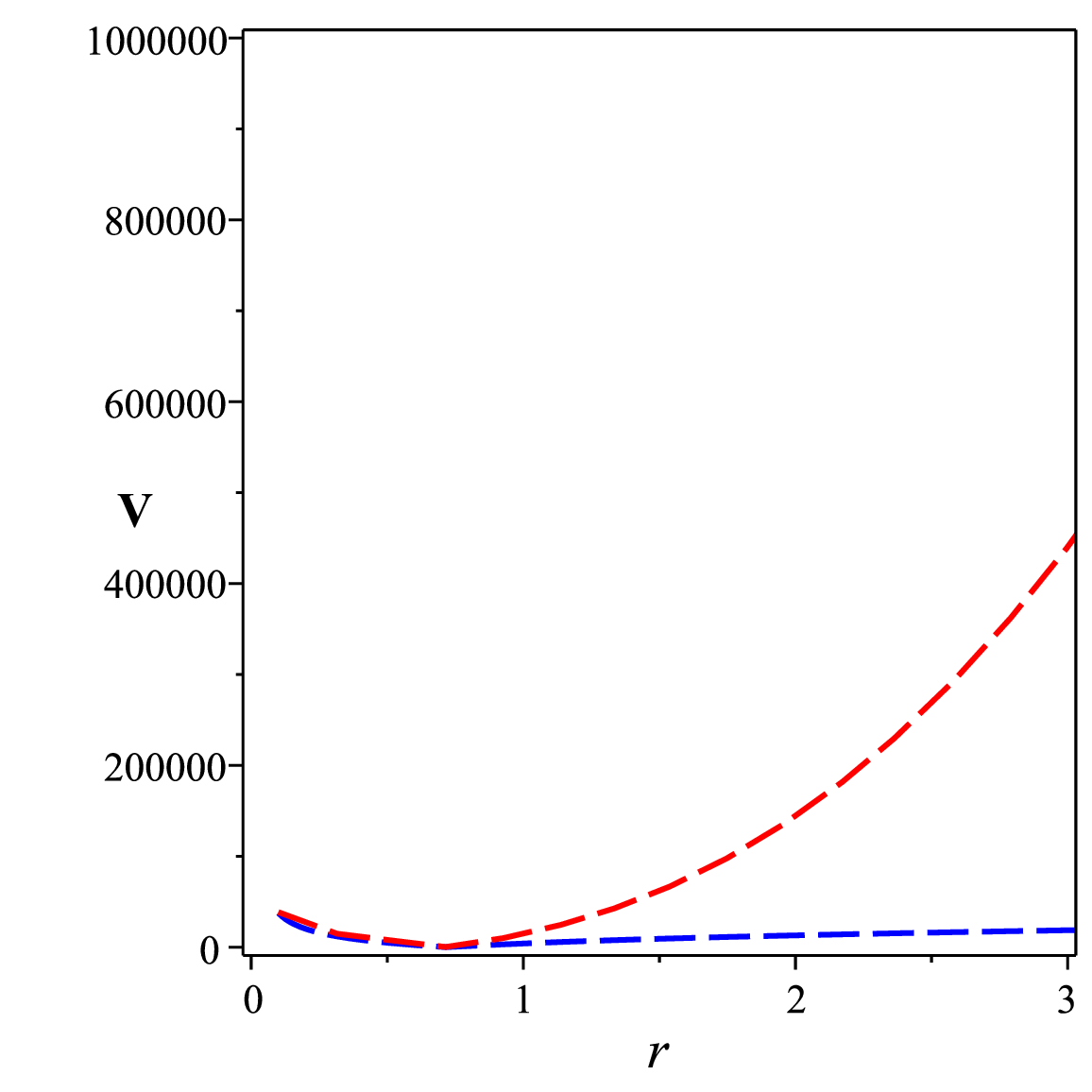}}
\subfigure[~The behavior of the second derivative of the potential associated with the solution given by  Eq. \eqref{ggsol}]{\label{fig:potddm}\includegraphics[scale=0.21]{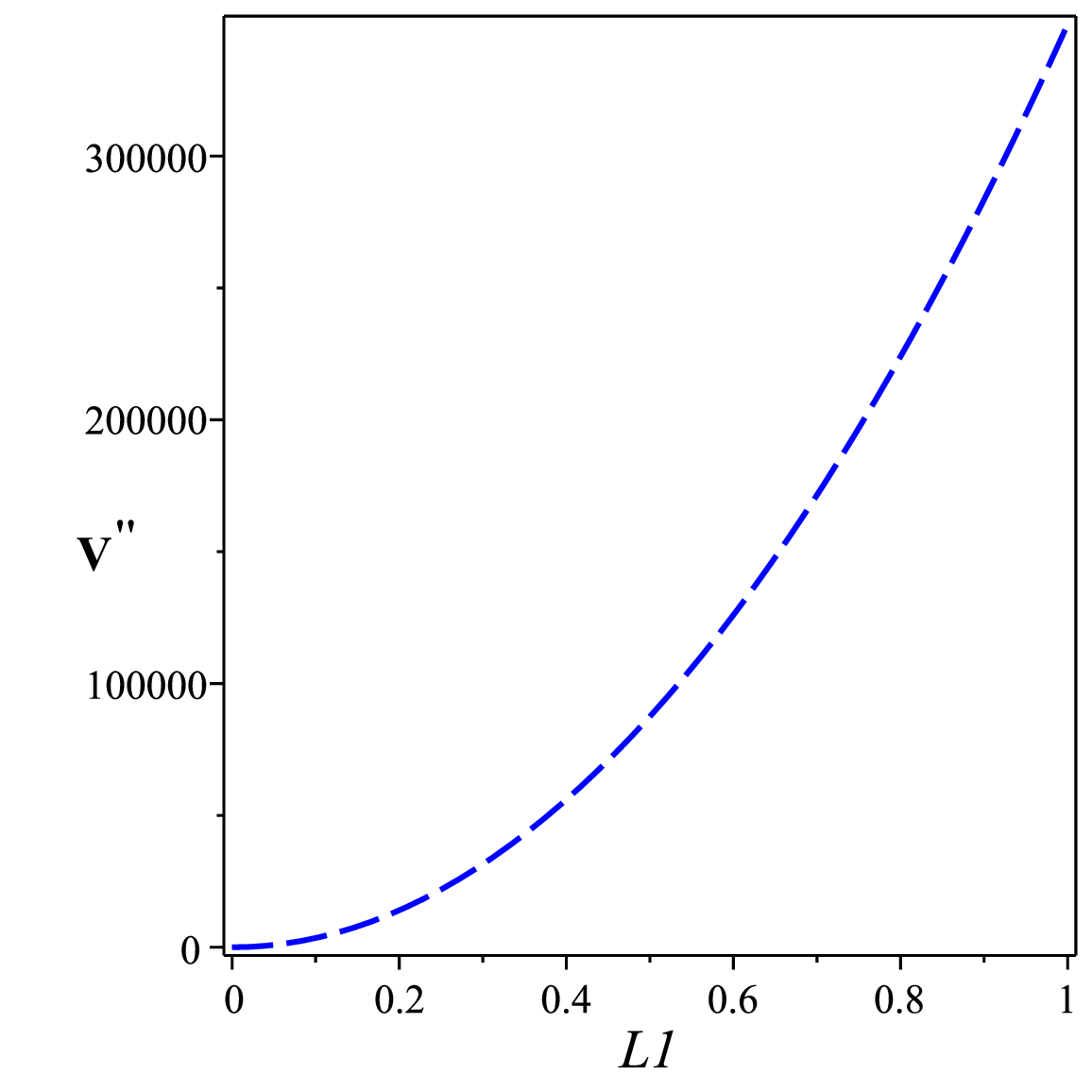}}
%
\caption{{ Multi-horizon black hole solution for the charged quadratic $f(\mathbb{Q})$
branch with $\alpha\neq1/(12\Lambda)$.
Panels \subref{fig:metm} and \subref{fig:metn} show the radial behavior of the metric functions $\mu(r)$ and $\nu(r)$
given by Eq.~(\ref{ggsol}), illustrating the presence of multiple horizons.
Panels \subref{fig:tempm}--\subref{fig:heatm} display the corresponding Hawking temperature, entropy, and heat capacity,
confirming thermodynamic consistency and stability of the solution.
Panels \subref{fig:potm} and \subref{fig:potddm}  show the second derivative of the effective potential,
demonstrating the stability of geodesic motion.
These plots illustrate a genuinely non-BTZ geometry arising from quadratic non-metricity
effects, with parameters $\Lambda=0.041$, $\alpha=-0.85$, $m=0.07$, $c_{2}=0.45$, and $c_{3}=0.6$.}}

\label{Fig:4}
\end{figure}

Now to check the viability of the above model we follow the procedure done in the pervious sections to calculate the thermodynamics and potential. To do this, we will plot the output of these calculations in Fig.~\ref{Fig:4}~\subref{fig:tempm}, ~\subref{fig:entm}, and ~\subref{fig:heatm}. Furthermore, we calculate the potential associated with the solution given by Eq.~\eqref{ggsol} and we plot it in panel Fig.~\ref{Fig:4}~\subref{fig:potm} and also calculate the second derivative of this potential and plot it in panel Fig.~\ref{Fig:4}~\subref{fig:potddm} which shows that the solution given by Eq.~ \eqref{ggsol} is a stable solution.

 \section{Discussion and Conclusions}\label{conclusion}

In this work, we presented new black hole solutions in (2+1)-dimensional spacetime within the framework of the quadratic form of symmetric teleparallel gravity, specifically $f(\mathbb{Q})=\mathbb{Q}+\alpha \mathbb{Q}^2-2\Lambda$, where $\mathbb{Q}$ is the non-metricity scalar. By introducing both charged and uncharged scenarios, we explored the effects of the quadratic correction term  on the geometry, thermodynamics, and stability of the  black holes.

Our results reveal several key findings. In particular, the  geometric deviations w.r.t. the General Relativity.
The uncharged solutions recover the BTZ black holes in the limit $\alpha \to 0$, consistent with General Relativity. However, when $\alpha\neq 0$, we obtain modified black holes whose curvature and non-metricity invariants deviate from the { BTZ} solutions. Interestingly, these deviations persist even asymptotically and suggest richer structure in the gravitational field when compared with the standard BTZ geometry.

{ Another important feature is the presence of weaker singularities. In fact,
the curvature and non-metricity invariants in the modified theory exhibit weaker singularities at $\alpha \to 0$  when $\alpha\neq 0$, with leading terms scaling as $\cal{O}\left(\frac{1}{r}\right)$, unlike the General Relativity counterparts where singular behavior scales as $\cal{O}\left(\frac{1}{r^2}\right)$. This demonstrates that the singularity is milder and has been rigorously characterized using the Tipler/KrÃ³lak \cite{Tipler:1977eb} criteria with explicit parameter ranges.}

This fact indicates a "softening" of the central singularity due to the non-metricity contributions, implying potentially regular or semi-regular spacetimes in modified gravity.

We performed a detailed thermodynamic analysis of the black hole solutions. The Hawking temperature, entropy, and heat capacity have been computed and shown to satisfy the first law of thermodynamics. The entropy is always positive, and the specific heat displays stable behavior in a large  parameter region, suggesting thermodynamic stability of the derived solutions.

Furthermore, we investigated the effective potential for both time-like and null geodesics  comparing the behavior of test particles in the modified theory and General Relativity. The quadratic correction   influences the location and stability of circular photon orbits. The second derivative of the effective potential confirms the presence of stable photon spheres under suitable conditions.

Finally, a notable feature of the general charged solution with $\alpha\neq \frac{1}{12\Lambda}$ is the presence of multiple horizons. These multi-horizon geometries enrich the causal structure and may allow new interpretations in thermodynamic and quantum gravity contexts.

In conclusion, the quadratic extension of $f(\mathbb{Q})$ gravity in lower-dimensional spacetime yields a diverse landscape of black hole solutions. These solutions not only generalize the well-known BTZ black holes but also introduce modifications in geometry, singularity structure, thermodynamic behavior, and orbital dynamics. Our analysis confirms that such modifications are physically meaningful and potentially observable in settings where non-metricity effects become significant. This study opens the door to further investigations of non-metricity-driven gravity in both theoretical and phenomenological directions, especially in the context of quantum gravity, holography, and gravitational wave physics.

Future extensions could include rotating or higher-dimensional analogs, as well as perturbative stability analysis in the presence of non-metricity. These directions would further clarify the physical implications of $f(\mathbb{Q})$  gravity in strong-field regimes.

{
We conclude by summarizing the genuinely new aspects of the solutions obtained in this work and clarifying their relation to the standard BTZ black holes and their extensions in modified gravity.

\begin{itemize}
    \item \textbf{Relation to standard (charged) BTZ solutions.}
    In the limit $\alpha = 0$ and $\nu_1(r)=1$, our solutions reduce exactly to the uncharged and charged BTZ black holes of Einstein--Maxwell gravity in $(2+1)$ dimensions.

    \item \textbf{New charged black hole family in quadratic $f(\mathbb{Q})$  gravity.}
    For $\alpha \neq 0$ and in the presence of an electric field, we obtain a genuinely new class of exact solutions in quadratic symmetric teleparallel gravity, characterized by a nontrivial radial deformation function $\nu_1(r)$. Although the electromagnetic potential preserves its logarithmic form, the spacetime geometry differs qualitatively from the charged BTZ solution and reduces to it smoothly only in the limit $\alpha \to 0$.

    \item \textbf{Modified asymptotic behavior.}
    The quadratic non-metricity corrections lead to asymptotics that are distinct from the BTZ case. In the generic charged branch ($\alpha \neq 1/(12\Lambda)$), the metric functions exhibit a leading $r^3$ behavior at large $r$, in contrast with the $r^2$ growth of BTZ spacetimes. In the special branch $\alpha = 1/(12\Lambda)$, curvature and non-metricity invariants acquire universal $1/r$ corrections, signaling long-range deviations from constant-curvature asymptotics.

    \item \textbf{Singularity structure.}
    The presence of quadratic non-metricity terms modifies the behavior of the invariants near the central region. Compared to the Einstein--Maxwell BTZ case, the resulting singularities are weaker in the sense that the leading divergences are softened in the relevant parameter regimes, according to standard criteria based on invariant behavior.

    \item \textbf{Multi-horizon configurations.}
    For appropriate choices of the integration constants and coupling parameters, the new charged solutions admit multiple horizons. This richer causal structure is absent in the standard charged BTZ geometry and arises here as a direct consequence of the quadratic non-metricity corrections.
\end{itemize}
 
These results demonstrate that quadratic $f(\mathbb{Q})$  gravity in $(2+1)$ dimensions supports black hole solutions that go beyond previously known BTZ and BTZ-like spacetimes, both geometrically and physically, while consistently reproducing the standard cases in the appropriate limits.
}
\appendix
\section{Appendix A: The forms of $\Psi_i$ appeared in Eq.~(\ref{met222}) and (\ref{met22}) and the asymptotic of thermodynamics quantities}
The forms of $\Psi_i$, $i=1\cdots4$ appeared  in Eq.~\eqref{met222} are given by
\begin{align}
&\Psi_1={\frac {81\sqrt {|\Lambda|}}{4}{c_2}^{13}{\Lambda}^{3}} \left[\frac{2\Lambda{c_2}^{2}}3 -\frac{m}{12}+{c_2}^{2}\Lambda\ln  \left( \frac{r}{r_0}\right) \right]  \left[ 3{c_2}^{6}{\Lambda} ^{3}  \ln  \left( \frac{r}{r_0}\right)  +4/3{\Lambda}^{ 2} \left( \Lambda{c_2}^{2}-\frac{3m}4 \right) {c_2}^{4}  \ln  \left( \frac{r}{r_0}\right)-{\frac {4}{ 27}}{\Lambda}^{3}{c_2}^{6}\right.\nonumber\\
  &\left.-\frac{m}9\Lambda{c_2}^{2} \left( \Lambda{c_2}^{2}-\frac{3m}{16} \right) \ln  \left( \frac{r}{r_0}\right) -{\frac {1}{1728}}{m}^{3}+{\frac {1}{216}}{c_2}^{2}{m}^{2}\Lambda \right]\,, \nonumber\\
   &\Psi_2=\frac {1}{256{ \Lambda}^{2}{c_2}^{10}}\left[ 2\left( 432\,m {c_2}^{4}{\Lambda}^{2}-2304\,{\Lambda}^{3}{c_2}^{6} \right) \ln \left( \frac{r}{r_0}\right)-5184\,{c_2}^{6}{\Lambda}^{3}  \ln \left( \frac{r}{r_0}\right)+ \left( -36\,{c_2}^{2}{m}^{2}\Lambda+384\,m{c_2}^{4}{\Lambda}^{2}-768\,{\Lambda}^{3}{c_2}^{6} \right)\right.\nonumber\\
  &\left. \ln \left( \frac{r}{r_0}\right) -16\,{c_2}^{2}{m}^{2}\Lambda+128\,{\Lambda}^{3}{c_2}^{6}+ 64\,m\,{c_2}^{4}{\Lambda}^{2}+{m}^{3}\right]\,,\quad \Psi_3=\frac {9 \left(\frac{2\Lambda\,{c_2} ^{2}}3-\frac{m}{12}+{c_2}^{2}\Lambda\,\ln \left( \frac{r}{r_0}\right)  \right) ^{2 } }{9{c_2}^{7}{-\Lambda}^{3/2}}, \nonumber\\
   &\Psi_4=\frac{1}{16}{\frac {16\,{\Lambda}\,{c_2}^{2}-m+12\,{c_2}^{2}{\Lambda}\,\ln \left( \frac{r}{r_0}\right) }{{c_2}^{4}{
\Lambda}}}. 
\end{align}
and the ones of $\Psi_i$, $i=5\cdots 8$ appeared in Eq.~\eqref{met22} are given by
\begin{align}
 &\Psi_5=-{\frac {9\sqrt {3}}{32}{\alpha}^{2}{c_2}^{6}} \left[ -12m\alpha  \ln  \left( \frac{r}{r_0}\right) -4 \left( {\frac {1}{108}}+{m}^{3}{\alpha}^{3} -3m{\alpha}^{2}\Lambda+ \left( \frac{1}4m+\frac{1}3\Lambda \right) \alpha \right) \ln \left( \frac{r}{r_0}\right) +{\frac {19}{648}}+{m}^{ 4}{\alpha}^{4}\right.\nonumber\\
 &\left.-6{m}^{2}{\alpha}^{3}\Lambda+ \left( {\frac {38} {9}}{\Lambda}^{2}+\frac{{m}^{2}}2+\frac{4m\Lambda}3 \right) { \alpha}^{2}+ \left( -{\frac {19}{27}}\Lambda+\frac{1}{27}m \right) \alpha+6 \left( {m}^{2}{\alpha}^{2}-\alpha\Lambda+\frac{1}6 \right) \ln  \left( \frac{r}{r_0}\right) +4\ln \left( \frac{r}{r_0}\right)\right] \sqrt {-\alpha{c_2}^{ 2}},\nonumber\\
 &\Psi_6={\frac {9}{16} {\alpha }{c_2}^{4}}\, \left[  \left(4\alpha\,\Lambda -\frac{1}3-3\,{m}^{2}{\alpha}^{2} \right) \ln  \left( \frac{r}{r_0}\right)- 3 \ln   \left( \frac{r}{r_0}\right) +{\frac {1}{54}}+{m}^{3}{\alpha}^{3}-4\,m{\alpha }^{2}\Lambda+ \left( \frac{m}3+\frac{2\Lambda}3 \right) \alpha+6 \,m\alpha\ln  \left( \frac{r}{r_0}\right) \right],\nonumber\\
  &\Psi_7={\frac {9}{32}}\,{\frac {\left[ -8/3\,\alpha\,{c_2}^{ 2}  \ln  \left( \frac{r}{r_0}\right)  -4/3\,{c_2}^{2} \alpha\, \left( 1/6-2\,\alpha\,\Lambda+{m}^{2}{\alpha}^{2} \right) +8/3\,{\alpha}^{2}{c_2}^{2}m\ln  \left( \frac{r}{r_0}\right) \right] \sqrt {-3\alpha\,{c_2}^{2}}}{{\alpha}^{2}{c_2}^ {6}}},\nonumber\\
   &\Psi_8={\frac {9}{16}}\,{\frac {-4/3\,{c_2}^{2}m{\alpha}^{2}+4/3\,
\alpha\,{c_2}^{2}\ln  \left( \frac{r}{r_0}\right) }{\alpha\,{c_2}^{4}}}
\,.\end{align}
The asymptotic form of Eq.~\eqref{ent1} takes the form
\begin{align}
&S(r_+)_{Eq.~\eqref{ent1}}\approx64\pi  \left[  \left(  \left\{  \left( 60{r}^{4}+36{r}^{6} \right) {\Lambda}^{3}-60{c_2}^{2}{r}^{2}{\Lambda }^{2}-36\Lambda{c_2}^{4} \right\} {\alpha}^{3}+ \left\{  \left( 3{r}^{6}+{r}^{4} \right) {\Lambda}^{2}+38{r} ^{2}\Lambda{c_2}^{2}+7{c_2}^{4} \right\} { \alpha}^{2}-\frac{1}4 \left\{  \left( 11{r}^{4} \Lambda\right.\right.\right.\right.\nonumber\\
&\left.\left.\left.\left.+5{r}^{6} \right)+11{r}^{2}{c_2}^{2} \right\} \alpha+\frac{1}{16}{r}^{4} \left( {r}^{2}+3 \right)  \right) \sqrt {1-12\alpha\Lambda }+ \left( 72{c_2}^{4}{\Lambda}^{2}+72{c_2}^{2} {r}^{2}{\Lambda}^{3}-144{r}^{4}{\Lambda}^{4} \right) { \alpha}^{4}+ \left\{  \left( 96{r}^{4}+144{r}^{6} \right) {\Lambda}^{3}\right.\right.\nonumber\\
&\left.\left.-234{c_2}^{2}{r}^{2}{\Lambda}^{2}-84\Lambda{c_2}^{4} \right\} {\alpha}^{3}+ \left\{  \left( 14 {r}^{4}+12{r}^{6} \right) {\Lambda}^{2}+{\frac {107}{2}}{r }^{2}\Lambda{c_2}^{2}+13/2{c_2}^{4} \right\} { \alpha}^{2}+ \left\{  \left( -5{r}^{4}-4{r}^{2}+{r}^{4} \left( -12 \alpha\Lambda+1 \right) ^{3/2} \right) \Lambda\right.\right.\nonumber\\
&\left.\left.-{\frac {23}{8}}{c_2}^{2} \right\} {r}^{2}\alpha+\frac{1}4 \left( \frac{3}4+ \left( 1+ \left( 1-12\alpha\Lambda \right) ^{3/2} \right) {r}^{2} \right) {r}^{4} \right] {\frac {1}{\sqrt {1-12\alpha\Lambda}{r}^{5} \left( 4\alpha\Lambda+1 \right) \left( 12\alpha\Lambda-1-2\sqrt {1-12\alpha\Lambda} \right) ^{2}}}\,.
\end{align}
The asymptotic form of Eq.~\eqref{heat1} takes the form:
\begin{align}
&H(r_+)_{Eq.~\eqref{heat1}}\approx -4\left[\left( 36{r}^{3}{|\alpha|}^{3/2}\Lambda+{r}^{3} \sqrt{|\alpha|}+36c_2\sqrt {3}{\alpha}^{2}{r}^{2}\Lambda-3c_2\sqrt {3}\alpha{r}^{2}+36|\alpha|^{3/2}{c_2}^{2}r+24{c_2}^{3}\sqrt {3}{\alpha}^{2} \right) \pi\right.\nonumber\\
&\left.  \left(138\alpha{c_2}^{2}{r} ^{2} -3{r}^{4}-520{\alpha}^{2}{c_2}^{4}+\sqrt {1-12\alpha\Lambda}[2880{\alpha} ^{3}{c_2}^{2}{r}^{2}{\Lambda}^{2}-1824{\alpha}^{2}{c_2}^{2}{r}^{2}\Lambda]-3456{\alpha}^{4}{c_2}^{2}{r}^{2}{\Lambda}^{3}-2568 {\alpha}^{2}{c_2}^{2}{r}^{2}\Lambda\right.\right.\nonumber\\
&\left.\left.+11232{\alpha}^{3} {c_2}^{2}{r}^{2}{\Lambda}^{2}+2880\sqrt {1-12\alpha\Lambda}{\alpha}^{3}\Lambda{c_2}^{4}-16{r}^{4} \sqrt {1-12\alpha\Lambda}{\alpha}^{2}{\Lambda}^{2}- 960{r}^{4}\sqrt {1-12\alpha\Lambda}{\alpha}^{3}{\Lambda}^{3}+44{r}^{4}\sqrt {1-12\alpha\Lambda}\alpha\Lambda\right.\right.\nonumber\\
&\left.\left.+132\alpha{c_2}^{2}{r}^{2}\sqrt {1-12\alpha\Lambda}+576\sqrt {1-12\alpha\Lambda}{r}^{6}{ \alpha}^{3}{\Lambda}^{3}-144\sqrt {1-12\alpha\Lambda}{r}^{6}{\alpha}^{2}{\Lambda}^{2}-52\sqrt {1-12\alpha\Lambda}{r}^{6}\alpha\Lambda+192{r}^{6}{\alpha}^{2}{\Lambda}^{2}\right.\right.\nonumber\\
&\left.\left.-80{r}^{6}\alpha\Lambda+5\sqrt {-12\alpha \Lambda+1}{r}^{6}-3{r}^{4}\sqrt {1-12\alpha\Lambda} -5760{\alpha}^{4}{c_2}^{4}{\Lambda}^{2}-1536{r}^{4}{ \alpha}^{3}{\Lambda}^{3}+2304{r}^{4}{\alpha}^{4}{\Lambda }^{4}-224{r}^{4}{\alpha}^{2}{\Lambda}^{2}+64{r}^{4}\alpha\Lambda\right.\right.\nonumber\\
&\left.\left.-560\sqrt {1-12\alpha\Lambda}{\alpha}^{2}{c_2}^{4}+2304{r}^{6}{\alpha}^{3}{\Lambda}^{3}+4{r}^{6} +6720{\alpha}^{3}\Lambda{c_2}^{4} \right) \right]\left\{{r}^{5} \sqrt {1-12\alpha\Lambda} \left( 4\alpha\Lambda+1 \right)  \left(1 -12\alpha\Lambda+2\sqrt {1-12\alpha\Lambda} \right) ^{2}\right.\nonumber\\
&\left.\left( -36{r}^{3}{\alpha}^{3/2}\Lambda-{r}^{3}\sqrt {\alpha}+36{\alpha}^{3/2}{c_2}^{2}r+48 i{c_2}^{3}\sqrt {3}{\alpha}^{2} \right) \right\}^{-1}\,.
\end{align}

The asymptotic form of the entropy for the case $\alpha=\frac{1}{12\Lambda}$:
\begin{align}
&S(r_+)
\approx64\pi  \left[  \left(  \left\{  \left( 60{r}^{4}+36{r}^{6} \right) {\Lambda}^{3}-60{c_2}^{2}{r}^{2}{\Lambda }^{2}-36\Lambda{c_2}^{4} \right\} {\alpha}^{3}+ \left\{  \left( 3{r}^{6}+{r}^{4} \right) {\Lambda}^{2}+38{r} ^{2}\Lambda{c_2}^{2}+7{c_2}^{4} \right\} { \alpha}^{2}-\frac{1}4 \left\{  \left( 11{r}^{4}+5{r}^{6} \right) \Lambda\right.\right.\right.\nonumber\\
&\left.\left.\left.+11{r}^{2}{c_2}^{2} \right\} \alpha+\frac{1}{16}{r}^{4} \left( {r}^{2}+3 \right)  \right) \sqrt {1-12\alpha\Lambda }+ \left( 72{c_2}^{4}{\Lambda}^{2}+72{c_2}^{2} {r}^{2}{\Lambda}^{3}-144{r}^{4}{\Lambda}^{4} \right) { \alpha}^{4}+ \left\{  \left( 96{r}^{4}+144{r}^{6} \right) {\Lambda}^{3}-234{c_2}^{2}{r}^{2}{\Lambda}^{2}\right.\right.\nonumber\\
&\left.\left.-84\Lambda{c_2}^{4} \right\} {\alpha}^{3}+ \left\{  \left( 14 {r}^{4}+12{r}^{6} \right) {\Lambda}^{2}+{\frac {107}{2}}{r }^{2}\Lambda{c_2}^{2}+13/2{c_2}^{4} \right\} { \alpha}^{2}+ \left\{  \left( -5{r}^{4}-4{r}^{2}+{r}^{4} \left( -12 \alpha\Lambda+1 \right) ^{3/2} \right) \Lambda-{\frac {23}{8}}{c_2}^{2} \right\} {r}^{2}\alpha\right.\nonumber\\
&\left.+1/4 \left( \frac{3}4+ \left( 1+ \left( 1-12\alpha\Lambda \right) ^{3/2} \right) {r}^{2} \right) {r}^{4} \right] {\frac {1}{\sqrt {1-12\alpha\Lambda}{r}^{5} \left( 4\alpha\Lambda+1 \right) \left( 12\alpha\Lambda-1-2\sqrt {1-12\alpha\Lambda} \right) ^{2}}}\,.\label{ent11}
\end{align}
The asymptotic form of heat capacity for the case $\alpha=\frac{1}{12\Lambda}$:
\begin{align}
&H(r_+)\approx -4\left[\left( 36{r}^{3}{|\alpha|}^{3/2}\Lambda+{r}^{3} \sqrt {|\alpha|}+36ic_2\sqrt {3}{\alpha}^{2}{r}^{2}\Lambda-3c_2\sqrt {3}\alpha{r}^{2}+36{|\alpha|}^{3/2}{c_2}^{2}r+24{c_2}^{3}\sqrt {3}{\alpha}^{2} \right) \pi\right.\nonumber\\
&\left.  \left(138\alpha{c_2}^{2}{r} ^{2} -3{r}^{4}-520{\alpha}^{2}{c_2}^{4}+\sqrt {1-12\alpha\Lambda}[2880{\alpha} ^{3}{c_2}^{2}{r}^{2}{\Lambda}^{2}-1824{\alpha}^{2}{c_2}^{2}{r}^{2}\Lambda]-3456{\alpha}^{4}{c_2}^{2}{r}^{2}{\Lambda}^{3}-2568 {\alpha}^{2}{c_2}^{2}{r}^{2}\Lambda\right.\right.\nonumber\\
&\left.\left.+11232{\alpha}^{3} {c_2}^{2}{r}^{2}{\Lambda}^{2}+2880\sqrt {1-12\alpha\Lambda}{\alpha}^{3}\Lambda{c_2}^{4}-16{r}^{4} \sqrt {1-12\alpha\Lambda}{\alpha}^{2}{\Lambda}^{2}- 960{r}^{4}\sqrt {1-12\alpha\Lambda}{\alpha}^{3}{\Lambda}^{3}+44{r}^{4}\sqrt {1-12\alpha\Lambda}\alpha\Lambda\right.\right.\nonumber\\
&\left.\left.+132\alpha{c_2}^{2}{r}^{2}\sqrt {1-12\alpha\Lambda}+576\sqrt {1-12\alpha\Lambda}{r}^{6}{ \alpha}^{3}{\Lambda}^{3}-144\sqrt {1-12\alpha\Lambda}{r}^{6}{\alpha}^{2}{\Lambda}^{2}-52\sqrt {1-12\alpha\Lambda}{r}^{6}\alpha\Lambda+192{r}^{6}{\alpha}^{2}{\Lambda}^{2}\right.\right.\nonumber\\
&\left.\left.-80{r}^{6}\alpha\Lambda+5\sqrt {-12\alpha \Lambda+1}{r}^{6}-3{r}^{4}\sqrt {1-12\alpha\Lambda} -5760{\alpha}^{4}{c_2}^{4}{\Lambda}^{2}-1536{r}^{4}{ \alpha}^{3}{\Lambda}^{3}+2304{r}^{4}{\alpha}^{4}{\Lambda }^{4}-224{r}^{4}{\alpha}^{2}{\Lambda}^{2}+64{r}^{4}\alpha\Lambda\right.\right.\nonumber\\
&\left.\left.-560\sqrt {1-12\alpha\Lambda}{\alpha}^{2}{c_2}^{4}+2304{r}^{6}{\alpha}^{3}{\Lambda}^{3}+4{r}^{6} +6720{\alpha}^{3}\Lambda{c_2}^{4} \right) \right]\left\{{r}^{5} \sqrt {1-12\alpha\Lambda} \left( 4\alpha\Lambda+1 \right)  \left(1 -12\alpha\Lambda+2\sqrt {1-12\alpha\Lambda} \right) ^{2}\right.\nonumber\\
&\left.\left( -36{r}^{3}{\alpha}^{3/2}\Lambda-{r}^{3}\sqrt {\alpha}+36{\alpha}^{3/2}{c_2}^{2}r+48 i{c_2}^{3}\sqrt {3}{\alpha}^{2} \right) \right\}^{-1}\,.\label{heat11}
\end{align}
\section*{Acknowledgements}
S.C.  acknowledges the Istituto Nazionale di Fisica Nucleare (INFN) Sez. di Napoli,  Iniziative Specifiche QGSKY and MoonLight-2  and the Istituto Nazionale di Alta Matematica (INdAM), gruppo GNFM, for the support. This paper is based upon work from COST Action CA21136 -- Addressing observational tensions in cosmology with systematics and fundamental physics (CosmoVerse), supported by COST (European Cooperation in Science and Technology).


\end{document}